\documentclass{aa}
\usepackage{graphicx,times}
\usepackage{natbib}
\bibpunct{(}{)}{;}{a}{}{,}
\newcommand{\aap}{A\&A\ }
\newcommand{\aaps}{A\&AS\ }
\newcommand{\pasp}{PASP\ }
\newcommand{\apj}{ApJ\ }
\newcommand{\apjl}{ApJ\ }
\newcommand{\apjs}{ApJS\ }
\newcommand{\mnras}{MNRAS\ }
\newcommand{\aj}{AJ\ }

\def\mrk{Mrk~279}
\def\ngc{NGC~5548}
\def\ch{{\it Chandra}}
\def\hst{HST-STIS}
\def\xmm{XMM-{\it Newton}}

\def\kms{km\,s$^{-1}$}
\def\Halpha{\ifmmode {\rm H}\alpha \else H$\alpha$\fi}
\def\Hbeta{\ifmmode {\rm H}\beta \else H$\beta$\fi}
\def\Hgamma{\ifmmode {\rm H}\gamma \else H$\gamma$\fi}
\def\Hdelta{\ifmmode {\rm H}\delta \else H$\delta$\fi}
\def\Lya{\ifmmode {\rm Ly}\alpha \else Ly$\alpha$\fi}
\def\Lyb{\ifmmode {\rm Ly}\beta \else Ly$\beta$\fi}
\def\Lyg{\ifmmode {\rm Ly}\beta \else Ly$\gamma$\fi}

\def\fexx{Fe\,{\sc xx}}

\def\heii{He\,{\sc ii}}
\def\ci{C\,{\sc i}}
\def\cii{C\,{\sc ii}}
\def\ciii{\ifmmode {\rm C}\,{\sc iii} \else C\,{\sc iii}\fi}
\def\civ{\ifmmode {\rm C}\,{\sc iv} \else C\,{\sc iv}\fi}
\def\cv{\ifmmode {\rm C}\,{\sc v} \else C\,{\sc v}\fi}
\def\cvi{\ifmmode {\rm C}\,{\sc vi} \else C\,{\sc vi}\fi}
\def\ni{N\,{\sc i}}
\def\nii{N\,{\sc ii}}
\def\niii{N\,{\sc iii}}

\def\nv{N\,{\sc v}}
\def\nvi{N\,{\sc vi}}
\def\nvii{N\,{\sc vii}}
\def\oi{O\,{\sc i}}
\def\oii{O\,{\sc ii}}
\def\oiii{O\,{\sc iii}}
\def\o5007{[O\,{\sc iii}]\,$\lambda5007$}
\def\oiv{O\,{\sc iv}}
\def\ov{O\,{\sc v}}
\def\ovi{O\,{\sc vi}}
\def\ovii{O\,{\sc vii}}
\def\oviii{O\,{\sc viii}}

\def\neix{Ne\,{\sc ix}}
\def\nex{Ne\,{\sc x}}

\def\mgxi{Mg\,{\sc xi}}
\def\siiv{Si\,{\sc iv}}
\def\siII{Si\,{\sc ii}}

\def\sixiii{Si\,{\sc xiii}}

\def\siv{S\,{\sc iv}}

\def\svi{S\,{\sc vi}}

\def\fexvii{Fe\,{\sc xvii}}

\def\fexix{Fe\,{\sc xix}}
\def\fexxiii{Fe\,{\sc xxiii}}
\def\fexxiv{Fe\,{\sc xxiv}}

\def\o{\o}
\def\ltsim{\raisebox{-.5ex}{$\;\stackrel{<}{\sim}\;$}}

\begin{document}

\title{X-Ray/Ultraviolet Observing Campaign of the Markarian 279 Active Galactic Nucleus Outflow:\\
a close look at the absorbing/emitting gas with \ch-LETGS }

\author{ E.~Costantini \inst{1,}\inst{2}
	\and
	J.S.~Kaastra \inst{1}
         \and
         N.~Arav \inst{3}
         \and
         G.A.~Kriss \inst{4}
         \and
         K.C.~Steenbrugge \inst{5}
         \and
         J.R.~Gabel \inst{3}
         \and
	 F.~Verbunt \inst{2}
	 \and
	 E.~Behar  \inst{6}
	 \and
	C.M.~Gaskell  \inst{7}
	 \and
	K.T.~Korista \inst{8}
	 \and
	D.~Proga \inst{9}
	 \and
	J.~Kim Quijano \inst{4}
	 \and
	J.E.~Scott \inst{10}
	\and
	E.S.~Klimek\inst{7}
	\and
	C.H.~Hedrick\inst{7}
	 }
\offprints{E. Costantini}
\mail{e.costantini@sron.nl}

\institute{ SRON National Institute for Space Research, 
              Sorbonnelaan 2, 3584 CA Utrecht, The Netherlands
              \and
		Astronomical Institute, Utrecht University, 
		P.O. Box 80000, 3508 TA, Utrecht, The Netherlands
              \and
		CASA, University of Colorado, 389 UCB, Boulder, CO 80309-0389, USA
             \and
              Space Telescope Science Institute, 3700 San Martin Drive, Baltimore, MD 21218, USA
             \and
             University of Oxford, St John's College Research Centre, Oxford, OX1 3JP, UK
	      \and
	      Department of Physics, Technion, Haifa 32000, Israel
		\and
		Department of Physics and Astronomy, University of Nebraska, Lincoln, NE 68588-0111	      
	      \and
	      Department of Physics, Western Michigan University, Kalamazoo, MI 49008
	      \and
	      Department of Physics, University of Nevada, 4505 South Maryland Parkway, Las Vegas, NV 89154
	      \and
	      Department of Physics, Astronomy, and Geosciences, Towson University, Towson, Maryland 21252 USA
	      	        }

\date{Received  / Accepted  }

\authorrunning{E.~Costantini et al.}
\titlerunning{\ch-LETGS observation of \mrk}

\abstract{We present a  \ch-LETGS observation of the Seyfert~1 galaxy \object{\mrk}. This observation 
was carried out simultaneously with \hst\ and FUSE, 
in the context of a multiwavelength study of this source. 
The \ch\ pointings were spread over ten days for a total exposure time of $\sim$360 ks.
The maximal continuum flux variation is of the order of 30\%.  
The spectrum of \mrk\ shows evidence of broad emission features, especially at the wavelength of the \ovii\ triplet. 
We quantitatively explore the 
possibility that this emission is produced  
in the broad line region (BLR). 
We modeled the broad UV emission lines seen in the FUSE and HST-STIS spectra 
following the ``locally optimally emitting cloud" 
approach. This method considers the emission from BLR as arising from ``clouds" with a wide range of
densities and distance from the source. We find that the X-ray lines luminosity derived from the 
best fit BLR model can match the X-ray features, suggesting that the gas producing the UV lines is sufficient to
account also for the X-ray emission. 
The spectrum is absorbed by ionized gas whose total column density is $\sim5\times10^{20}$\,cm$^{-2}$. 
The absorption spectrum can be modeled by two distinct gas components (log$\xi\sim 0.47$ and $2.49$, respectively)
both showing a significant outflow velocity. However, the data allow also the presence of intermediate ionization components. 
The distribution of the column densities of such extra components as a function of the ionization parameter 
is not consistent with a continuous, power law-like, absorber, suggesting a 
complex structure for the gas outflow for \mrk.     
\keywords{Galaxies: individual: Mrk~279 -- Galaxies: Seyfert -- quasars: absorption lines -- quasars: emission lines --
 X-rays: galaxies }}
\maketitle

\section{Introduction}

Active galactic nuclei (AGN) are believed to be powered by accretion into a super massive black hole (BH).   
Through the identification of narrow spectral features in the energy range $E<$ 2\,keV, 
the basic phenomenology of the absorbing ionized gas out-flowing from the vicinity of the BH 
has been established: approximately 60\% of the Seyfert~1 galaxies display such an outflow \citep{crenshaw99}. 
 Objects with an UV absorber always display an X-ray absorption component \citep{crenshaw99,kriss02}. 
Recent results show that the absorber has to be a multi-ionization component gas and some 
of the components of the UV and X-ray outflow may be part of the same material \citep{gabel05a}. 
Only in the X-ray band higher ionized components, producing mostly He-like and H-like ions of C, N, O, 
and up to H-like iron, can be detected. These high ionization components have been reported to have 
a larger column density \citep[e.g., ][]{turner,netzer03} and sometimes a higher outflow velocity
\citep[e.g.,][]{steen05}. 
The physical structure of the absorbing systems related to the BH activity is not understood. 
Independent measurement of the physical 
structure of the out-flowing ionized gas led to incompatible results. 
The density structure may be continuous as a function
of the ionization level of the gas 
(NGC~5548, Kaastra et al. 2002, Steenbrugge et al. 2003). Alternatively, 
the absorber is clumped in discrete components, in pressure 
equilibrium surrounded by a hotter and tenuous phase \citep{kriss95,kriss01}. The spectrum of 
\object{NGC~3783} seems to be well described by this model \citep[][]{k03,netzer03}. 
Finally, there are conflicting results on the spatial extent and 
the distance from the ionizing source of such an absorber. 
\citet{behar03}, based on the \xmm\ observation of NGC~3783, 
proposed that the out-flowing X-ray absorber has the same spatial 
distribution as the narrow line region (NLR) mapped in UV, as confirmed by the 
UV spectral analysis of this source \citep{gabel05a}. This picture would be analogous to the geometry drawn for the 
Seyfert~2 galaxies (e.g. \object{NGC~1068}, \cite{kink02}; \object{Mrk~3}, \cite{sako00}).
Conversely, a \ch-LETGS and HETGS study on NGC~5548 suggested that 
the outflow is more collimated with a small 
opening angle \citep{steen05}. The location of the warm absorbers may be as far as the NLR, 
or closer to the source, at
$\sim$pc scale, where a molecular torus should exist \citep[see][ and references therein]{blustin} 
or, finally, the ionized gas outflow may be generated by accretion disk instabilities \citep[][]{proga}.
The X-ray line emission spectrum turned out to be an important component in Seyfert~1 galaxies.
Recombination lines, narrow and variable on long time scales 
are generally believed to arise in the NLR \citep[e.g., ][]{pounds}.
Moreover, broad emission features, several \AA\ wide, have recently been detected in the 
X-ray spectra of some Seyfert
galaxies. Whether 
some of them are due to curvatures of the primary continuum hiding absorption \citep[][]{lee}, or are 
relativistically broadened lines \citep{branduardi,sako,ogle04}, is still unclear.   
The modeling of other detected 
broad excesses as broad lines 
or blends of lines, in the soft X-ray spectrum, points to emission from the BLR 
\citep{steen05,ogle04}. A deeper understanding of the physics behind such emission features is also important for a correct
interpretation of the entire X-ray spectrum.

\mrk\ is a Seyfert~1.5 galaxy, located at redshift 
$z$\,=\,0.0305 \citep[][]{scott04}. 
The source spectrum is affected by a relatively low Galactic absorption 
\citep[$N_{\rm H}=1.64\times 10^{20}$~cm$^{-2}$, ][]{elvis89}.  
\mrk\ has been extensively
studied by low resolution instruments \citep[e.g.,][]{1995ApJ...447..121W,2001ApJ...550..261W} 
mainly by virtue of the 6.4 keV iron emission
line, which is characterized by a broad profile and a variable flux.
The first systematic study 
on the soft X-ray energy band was carried out in 2002 using \ch-HETGS, simultaneously with FUSE and HST-STIS \citep{scott04}. 
At that epoch both UV and X-ray observations show that the source was at 
the lower end of its historical flux range \citep{2001ApJ...550..261W}. The $2-10$ keV flux was 
$\sim1.2\times10^{-11}\ {\rm erg\ cm^{-2} s^{-1}}$ and the 1000\,\AA\ flux 
was 
$\sim0.13\times10^{-11}\ {\rm erg\ cm^{-2} s^{-1}}$\AA$^{-1}$. While monitored 
by FUSE, the source became 7.5 times dimmer from 1999 to 2002 and a similar
behavior was recorded in the
optical \citep{bachev04}. In the 2002 multiwavelength observation, a relatively short exposure time combined with a low flux state conspired against the detection of 
any significant absorption in the HETGS spectrum. 
The ionized absorber in \mrk\ has indeed quite a low column density
in the X-ray domain (${\rm N_H}\sim 5\times 10^{20} {\rm cm}^{-2}$, this study), as compared for example to NGC~3783
\citep[$\sim3\times10^{22}$ cm$^{-2}$,][]{kaspi01} or NGC~5548 \citep[$\sim5\times10^{21}$ cm$^{-2}$,][]{kaastra02}.\\ 
A large scale project has more recently focused on \mrk. 
The observations were carried out using HST-STIS, FUSE and {\it Chandra}-LETGS 
simultaneously. The methodology in studying the UV absorption is described in \citet[][hereinafter G05]{gabel05}, 
while \citet{arav05} 
focuses on the modeling of the structure of such absorber. 
For the LETGS data,  
the theoretical background for the line production from a meta-stable level of \ov\ is described in \citet{jelle04}, while
in the present paper we present the total modeling of the X-ray spectrum.\\ 
The X-ray flux we measure with \ch-LETGS is $\sim2.18\times10^{-11}\ {\rm erg\ cm^{-2} s^{-1}}$ 
in the $2-10$\,keV range, while the UV flux 
at 1000\,\AA\ is almost a factor of ten higher 
($\sim1.23\times10^{-11}\  {\rm erg\ cm^{-2} s^{-1}}$\AA$^{-1}$, G05) than the 2002 UV observation. 
The flux history in the optical V band, recorded by the 0.9-m University
of Nebraska telescope, shows that the source underwent a large flux rise 
during the $\sim$2 months preceding the present campaign (Gaskell et al. 2006, in prep.).\\

The paper is organized as follows. Sect.~2 is devoted to the analysis of the data and the modeling of the 
emission and absorption spectral features of \mrk. In Sect.~3 we discuss our results and in
Sect.~4 we present our conclusions. 
The cosmological parameters used are: 
H$_0$=70 km/s/Mpc, $\Omega_{\rm m}$=0.3, and $\Omega_{\Lambda}$=0.7.
The quoted errors refer to 84\% confidence ($\Delta\chi^2=2$ for one interesting parameter), unless otherwise stated.

\section{The data\label{sect:data}}

\begin{figure}
\resizebox{\hsize}{!}{\includegraphics[angle=90]{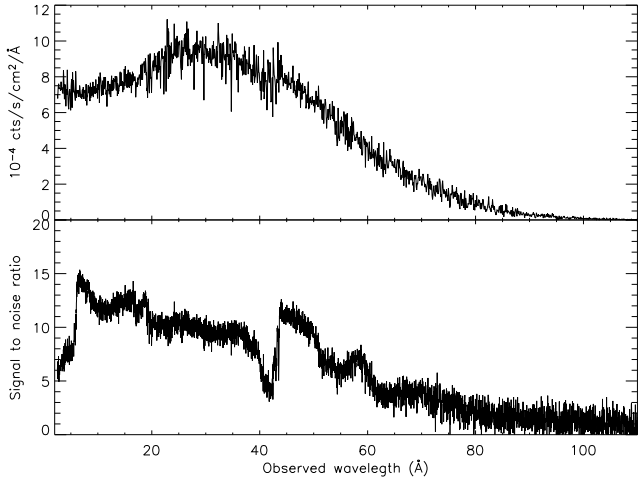}}
\caption{Upper panel: broad band fluxed spectrum of \mrk. The data have been binned by a factor of two. 
Lower panel: signal to noise ratio for the unbinned spectrum.}
\label{f:sn}
\end{figure}

\mrk\ was observed seven times spread over ten days for a total duration of 360 ks, 
by the Low Energy Transmission Grating Spectrometer \citep[LETGS,][]{2000ApJ...530L.111B} coupled with the HRC-S on board of \ch. The nominal 
wavelength range in which LETGS operates is 1.7$-$150 \AA, with an energy resolution of 0.05 \AA.  
In Tab.~\ref{t:obs_log} the observation log for \mrk\ is shown. 
The data reduction was carried out using the standard CXC pipeline up to the creation of 
 the level 1.5 event files. The steps leading to the final event file (level~2) 
 follow the same line as CXC for the wavelength accuracy
 determination and effective area, only make use of an independent procedure, first described in \citet{kaastra02}. 
 Higher spectral orders have been taken into account in all spectral fitting, while they were
 subtracted, using a bootstrap method, from the ``fluxed" spectrum (cts s$^{-1}$ cm$^{-2}$ \AA$^{-1}$) uniquely 
 for plotting purposes. 
The spectrum was differently re-binned, in channel space, depending on the features we wanted to focus on. 
A large re-binning (factor of 10) was used to model the continuum shape, while a binning of 2 was used for the study of the
spectral features. In Fig.~\ref{f:sn} (upper panel) the fluxed broad band spectrum is displayed, while 
the S/N of the unbinned spectrum is
shown in the lower panel.  
We have a S/N of at least 5 for all wavelengths between 2--58~\AA, except for a small region around the instrumental \ion{C}{i}
edge near 40~\AA.  Over most of the 6--48~\AA\ band, S/N is $\ga 10$.  This
limit of S/N=10 is often seen as a necessary condition for reliable
spectroscopy. The spectral analysis was carried out using the fitting package 
SPEX\footnote{http://www.sron.nl/divisions/hea/spex/version2.0/release/index.html} (ver.~2.0).

\begin{table}
\caption{\label{t:obs_log}\ch-LETGS observation log of \mrk.}
\begin{center}
\begin{tabular}{lll}
\hline
\hline
ObsId & Exp  & date\\
	& (ks) & dd-mm-yy\\
\hline
4401 &	55	&	$10-05-03$	\\
4027 &	100 	& 	$12-05-03$	\\
4400 &  50  	&	$14-05-03$	\\
4428 &	50	&	$15-05-03$	\\
4429 &	50	&	$17-05-03$	\\
4427 &	27.5	&	$19-05-03$	\\
4436 &	27.5	&	$20-05-03$	\\
\hline
\end{tabular}
\end{center}
\end{table}

\subsection{Time variability\label{sect:time}}

\begin{figure}
\resizebox{\hsize}{!}{\includegraphics[angle=90]{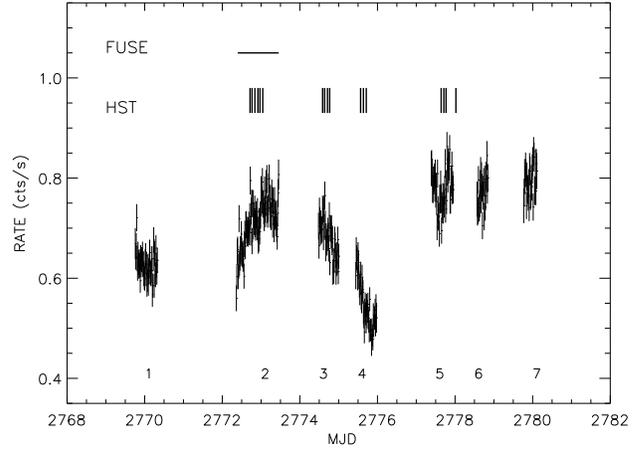}}
\caption{LETGS light curve of Mrk~279 in the zeroth spectral order. Bin size: 1000~s. 
The individual 7 time segments are labeled at the bottom. The FUSE and HST pointings are also marked.}
\label{fig:lc}
\end{figure}

The light curve of Mrk~279 taken in the zeroth spectral order is shown in
Fig.~\ref{fig:lc} where we mark also the epochs of the FUSE and HST pointings, during the multi-wavelength campaign. 
The count rate $c(t)$ of the X-ray data varies gradually, spanning the range
between 0.49~c/s (during the minimum in observation 4) to 0.82~c/s (at the
maximum of observations 5 and 7).   
No significant variations on time scales
shorter than 10~ks are observed.  Typical time scales ${\mathrm d} t/{\mathrm
d}\ln c(t)$ for the more gradual variations are $\sim$120~ks. Details on the X-ray variability will be
discussed elsewhere (Gaskell et al. 2006, in prep.).


\subsection{Continuum spectrum\label{sect:cont}}

\begin{figure}
\resizebox{\hsize}{!}{\includegraphics[angle=90]{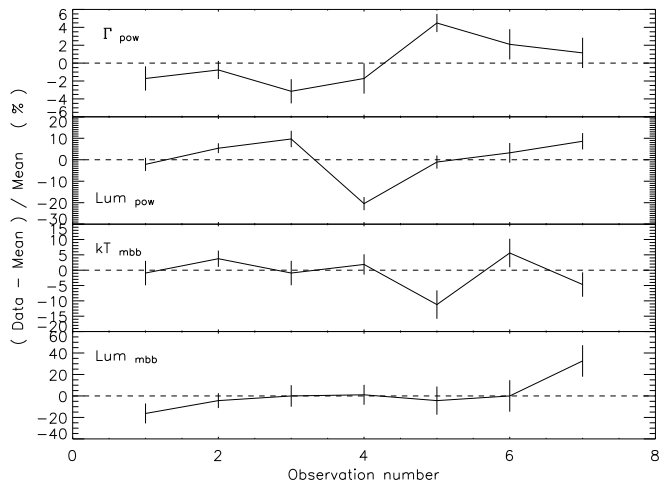}}
\caption{Percentage deviation from the parameter values of the combined spectrum (Tab.~\ref{tab:contpar}) for: the power law slope,
$\Gamma$, the power law 2--10 keV luminosity, the temperature (kT) and the luminosity of the modified black body. The
horizontal axis
labels the observation number as defined in Fig.~\ref{fig:lc}.}
\label{f:chi_cont}

\end{figure}

The continuum shape of the average spectrum is modeled by a power-law (average 
photon index $\Gamma=2.01\pm 0.01$), 
dominating at shorter wavelengths, and a
``modified" black body emission with an average temperature $kT=0.108\pm 0.002\,{\rm keV}$, 
to account for a ``soft excess" at longer wavelengths (18--80~\AA, Tab.\ref{tab:contpar}).   
The black body model takes into account the modifications by coherent Compton scattering \citep[see ][]{kaastrabarr}.  
The parameters of this fit are reported in Tab.~\ref{tab:contpar}. 
We have also fitted the spectrum of each of the seven observations using the same power law
plus modified blackbody spectrum as used above for the
combined spectrum. 
In Fig.~\ref{f:chi_cont} these continuum parameters  are plotted 
in terms of percentage deviation from the value found for the combined spectrum, as a function of the observation number 
(and therefore of time, Fig.~\ref{fig:lc}).  
We see that the power-law parameters agree within $\sim 2\sigma$ with the average value apart from two outlying 
points: the value of $\Gamma$ for observation 5 and the 2$-$10 keV luminosity for observation 4, to which the flux drop in
the light curve is ascribed. The modified black body parameters are more stable over time, the maximum deviation being $\sim 2\sigma$
from the mean. The stable shape of the soft thermal component dominates the band where most of the absorption and emission spectral features are, while the power
law dominates at 2--8 \AA, where the spectral resolution starts to degrade, making the (few) spectral features at those
short wavelength more
difficult to study. The largely stable shape of the spectrum supports the
assumption that the time-averaged continuum of the combined data can be used
in our modeling of the absorption/emission components.

\begin{table}

\caption{\label{tab:contpar}Fit parameters of a power law plus modified blackbody fit
to the total spectrum of \mrk. The errors on the parameters are only statistical and 
do not include any systematic uncertainty. The fit was restricted to
the 1.9--110~\AA\ range. The spectra were binned by a factor of 10 before
spectral fitting.}
\begin{center}
\begin{tabular}{llll}
\hline
\hline
$\Gamma$ & $L_{\mathrm{2-10\,keV}}$ & $T_{\mathrm{mbb}}$ & $L_{\mathrm{mbb}}$  \\
	&$10^{43}$ erg s$^{-1}$	& keV		&  $10^{43}$ erg s$^{-1}$  \\
\hline
2.01$\pm$0.01 &  9.3$\pm$0.2 & 0.108$\pm$0.002 &  9.2$\pm$0.5  \\
\hline
\end{tabular}
\end{center}
\end{table}

In the following we will describe first the modeling of the emission spectral features (broad and narrow). We will see how
this is a necessary step in order to further describe the absorbed spectrum.


\subsection{Emission lines}

\begin{figure}
\resizebox{\hsize}{!}{\includegraphics[angle=90]{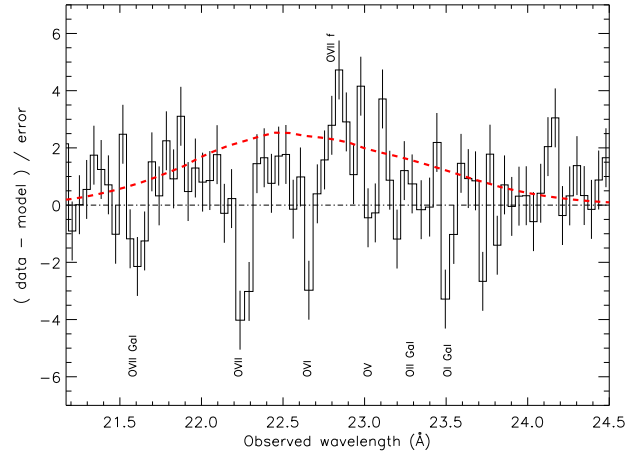}}
\caption{\label{f:res_po} 
Residuals, in terms of sigma, to the continuum model in the \ovii\ triplet region. A broad Gaussian (dashed line) 
has been added only to guide the eye. 
The continuum is represented by a power law plus a modified black body absorbed by galactic absorption. 
The labels refer to
the position of expected absorption/emission narrow lines.}
\end{figure}

Residuals to the continuum 
model show many narrow absorption and emission features and
also some broad excesses, roughly 1\,\AA\ wide, 
especially around the transitions of the most prominent oxygen ions in the X-ray 
spectrum (\ovii\ and \oviii).  
This can be seen, for instance, 
in Fig.~\ref{f:res_po}, where the region around the \ovii\ triplet is shown. 
We overplot the profile of a broad Gaussian line, only to guide the eye. 
Especially in this region, where any absorber is expected to show many (oxygen) features, 
the fitting of the absorption lines would 
be seriously compromised (e.g. \ov\ and the \oii\ Galactic absorption, Fig.~\ref{f:res_po}) if these broad
emission features are ignored. These excesses are present in all seven observations.   

The uncertainty associated with the flux/shape of the broad X-ray lines is such that we would not be able to 
quantify if the broad excesses
respond to the average flux variation of $\sim$25\% experienced by the central source. 
Therefore we considered the whole data set.\\


\subsubsection{A basic fit for the broad lines}\label{par:gauss}

We first attempted a purely phenomenological approach: we included broad Gaussian profiles, at the position of the main
H-like and He-like ions of C, N, O
to the continuum described in
Sect.~\ref{sect:cont}. The intrinsic flux of the lines is of course 
intercepted by the absorbers on the line of sight (Sect.~\ref{par:abs}), 
which must be consistently taken into account even in this
simple modeling. The resulting unabsorbed profile of the lines is thus expected to be significantly modified 
with respect to that qualitatively drawn in
Fig.~\ref{f:res_po}. 
A contribution by narrow emission lines could be included only for \ovii\ and \oviii, 
as they produce the only measurable narrow
recombination lines (Sect.~\ref{par:narrow}). For 
the \ovii\ triplet we also formally included, beside the measured narrow forbidden line, the recombination and 
intercombination lines with flux $1/4$ of the forbidden line \citep[assuming a purely photoionized gas with density
$<10^{9}$\,cm$^{-3}$,][]{delphine}.\\  
The centroid of the broad Gaussian lines are kept fixed to the 
 wavelength of the \Lya\ transition for \cvi, \nvii\ and \oviii. 
 For the blended lines, the centroid was at first left 
 free to vary in the range among the nominal lines' wavelength and later fixed to that value (Tab.~\ref{t:x_cloudy}). 
The width of the lines is a difficult parameter to determine. At this stage, we 
tentatively fixed the FWHM of the individual lines at $\sim$10000\,\kms. The width for a triplet is therefore a blend of individual
lines. 
For \ovii\ we actually fitted the full-width-half-maximum (FWHM) of the blended triplet, keeping it fixed later on to evaluate the errors on the line luminosity. 
In Tab.~\ref{t:gaussian} the result of this fit is shown.   
The intrinsic line luminosities listed in Tab.~\ref{t:gaussian} are derived considering a self-consistent
emission/absorption fit and the relative errors are evaluated from this best fit. 
In the last column we list the significance of
each broad excess individually, in terms of $\Delta\chi^2$, starting from a model with no broad lines. 
$\Delta\nu$ is 1 (where $\nu$
is the number of degrees of freedom) as the width and the centroid of the line are kept fixed. With those constraints, 
we determine that the \ovii\ triplet and the \oviii\ \Lya\ have a significant detection (approximatively 6 and 3 
$\sigma$, respectively).
Measuring the physical
parameters, such as flux and FWHM, from this line-by-line fitting might be a quantitative way to model these excesses. 
However, given their relative weakness, this approach leads to severe uncertainties. 
First, the absorption lines that are superimposed on
these profiles, the level of the continuum, and possible emission line blending, 
contribute to make the flux and width determination of these lines very uncertain. 
Second, the weaker lines, whose flux is only a few per cent higher than the continuum, can be easily missed.  

\begin{table}
\renewcommand{\tabcolsep}{1mm}
\caption{\label{t:gaussian} Parameters of the X-ray broad lines from a line-by-line fitting. The line transition, its
centroid (in \AA), the FWHM in \kms\ and the rounded correspondent value in \AA, the intrinsic line luminosity 
and the significance are listed (see text for details). Note that for the triplets the FWHM is the result of the blend of the lines.}
\begin{center}
\begin{tabular}{llllll}
\hline\hline
{\footnotesize Ion} & wavelength & FWHM & FWHM & Lum$_{\rm obs}$ & $\Delta\chi^2/\Delta\nu$ \\
&    \AA  & $10^{4}$\kms & \AA& ($10^{41}$erg s$^{-1}$) & \\
\hline
\cv$^{1,2}$ 	& 41.47 & 1.0 		     & 1.4			& $<3$  	& 3.4/1 \\ 
\cvi\ 		& 33.73 & 1.0		     & 1.1			& 1.1$\pm$0.5	& 7.1/1 \\
\nvi$^{2}$  	& 28.75 & 1.7 		     & 1.7			& $<1.1$	& 0.5/1 \\
\nvii\  	& 24.78 & 1.0 		     & 0.8			& $<1.7$	& 0.4/1 \\
\ovii$^{2}$  	& 21.90 & 1.9$^{+0.7}_{-0.4}$& 1.4$^{+0.5}_{-0.3}$	& 5.0$\pm$0.8	& 45/1  \\
\oviii$^3$  	& 18.96 & 1.6 		     & 1.0			& 2.4$\pm$1.0	& 13/1  \\ 
\neix$^{2}$ 	& 13.67 & 1.5 		     & 0.7			& $<2.5$	& 6.7/1 \\ 
\hline
\end{tabular}
\end{center}
{\footnotesize $^1$ In the instrumental \ci\ edge, the wavelength and the width refer the the forbidden line.\\
$^2$ Triplets.\\ 
$^3$ Blended with the \ovii\ He$\beta$.}
\end{table}


\subsubsection{A synthetic model for the broad lines:\\ the LOC model}\label{par:broad_data}

We looked for a physical model that would account for all the lines simultaneously. 
For instance, in analogy with the UV broad emission lines, 
one plausible possibility is that these emission features arise within the BLR.   
In order to 
self consistently test this idea, we first modeled the UV broad lines detected simultaneously by \hst\ and FUSE. 
From the best fit synthetic model we then infer the X-ray line luminosities from the same emission-line gas 
to be compared with the LETGS data.\\ 
%
%
In the modeling, we considered the luminosity, corrected for the galaxy extinction (E(B-V)=0.016), of the broad
component of the UV line profiles, listed in Tab.~\ref{t:broad_lines} 
(see also Fig.~1 of G05). The FWHM of the UV lines lie in the range $\sim8500-9500$\,\kms.
A full description of the UV data analysis of
the broad lines will be described in Scott et al. (2006, in prep.). 
%
%
\begin{figure}
\resizebox{\hsize}{!}{\includegraphics[angle=90]{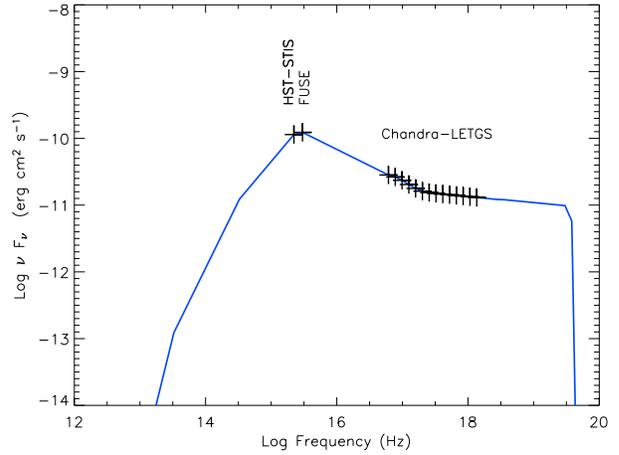}}
\caption{\label{f:sed}The spectral energy distribution for \mrk. The points at 1000 and 
1350\,\AA\ are the continuum flux of the simultaneous FUSE and \hst\ observation, respectively.
The X-ray continuum is taken from the present LETGS data.}
\end{figure}
We used Cloudy \citep{fer98}, ver.~95.06, to reproduce the UV emission from the BLR. 
The spectral energy distribution (SED) of the incident continuum that we used is shown in Fig.~\ref{f:sed}. 
The UV points at 1000 and 1350\,\AA\ come from the FUSE and HST-STIS measurements, respectively (G05),
while the X-ray ionizing
continuum comes from the time averaged LETGS data. 
Lacking any information of the high energy spectrum, we artificially cut off the X-ray 
power law at $\sim$150 keV. 
The low energy part of the SED resembles the shape assumed for the standard photoionizing continuum used in
Cloudy\footnote{ftp://gradj.pa.uky.edu/gary/cloudy\_gold/docs/hazy1\_06\_01\_rc1.pdf}.  
We kept the luminosity of the ionizing radiation fixed: $\log L=44.57$ erg\,s$^{-1}$, between 1 and 1000 Rydberg, 
as measured from the SED.\\ 
In order to reproduce emission from a large range of ionization stages, we followed the 
``locally optimally emitting clouds" prescription \citep[LOC, ][]{baldwin95}.
For this purpose, we created a grid of values for 
the density $n$ and the
distance $r$ of the BLR clouds. Once these parameters are set, the
ionization parameter is readily calculated: $\xi$=$L/nr^2$.   
We computed the integrated luminosity of the lines, 
weighted by a power law distribution of the density $n$ and the distance $r$ and assuming spherical symmetry \citep{baldwin95}:

\begin{equation}
L_{\rm line}\propto\int\int L(r,n)\ r^{\gamma}\ n^{\beta}\ dr\ dn.
\end{equation}

\begin{figure}

\begin{center}
\resizebox{\hsize}{!}{\includegraphics[angle=90]{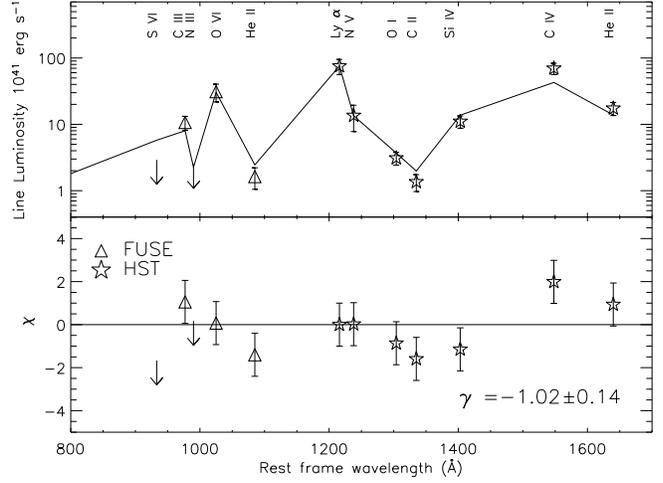}}

\end{center}
\caption{\label{f:cloudy_fit1} Best fit of the intrinsic luminosity of the UV broad emission lines (upper panel). The model
is displayed by a continuous line only to guide the eye. Lower panel: residuals in terms of $\sigma$ to the best fit.  
}
\end{figure}

\begin{figure}
\resizebox{\hsize}{!}{\includegraphics[angle=90]{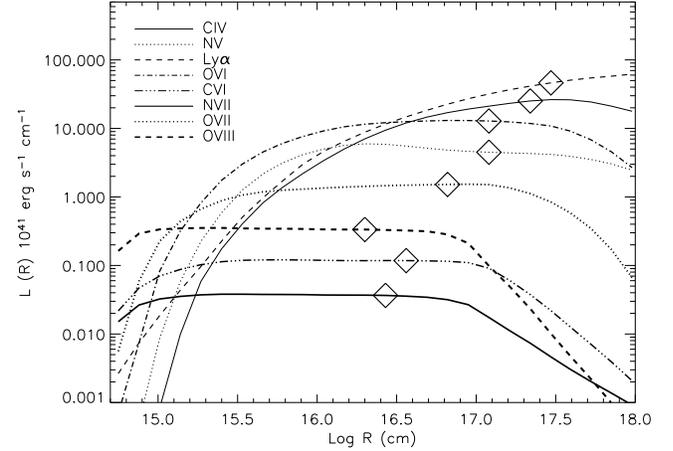}}
\caption{\label{f:prof_r_n} Best fit luminosity 
profiles of eight of the ions considered for the fit. The luminosity was
integrated at each radius using the weighting function $n^{\beta}$. 
For each emission line the diamonds indicate the luminosity-weighted radius. Note that the model underpredicts the observed
luminosity of \civ\ by $\sim$40\%.}
\end{figure}

The $\beta$ parameter was fixed to $-1$ \citep[e.g. ][]{korista00}, 
while the slope of $r$, $\gamma$, was a free 
parameter in the fit. 
The density $n$ ranged between $\sim$10$^{8-12.5}$ cm$^{-3}$ with a step of 0.125 in log. 
The same step was taken for the radius, which
ranged between 10$^{14.7-18}$ cm. Densities lower than $n=10^{8}$\,cm$^{-3}$ are not expected within the BLR, 
while for densities $n=10^{13-14}$\,cm$^{-3}$ the cloud emission would be, in most cases, 
continuum rather than lines \citep{kor97}. The wide range of distances allows us to completely follow the luminosity
distribution of the main X-ray lines as a function of radius. The gas located at distances 
$\ltsim 10^{15.3}$\,cm gives, for all lines, a minor contribution to the integrated luminosity. 
The column density was fixed to 10$^{23}$ cm$^{-2}$, as most of the 
emission should occur for $N_{\rm H}=10^{22-23}$\,cm$^{-2}$. Since 
in the observed line luminosity a main reason of uncertainty is, at all wavelengths, line blending, 
to bypass the line-blending problem, the lines predicted by Cloudy for each $n$ and $r$, 
which could be a result of a blend, were summed 
up (Tab.~\ref{t:broad_lines}).
The best fit was reached through a $\chi^2$ minimization.\\ 
In Fig~\ref{f:cloudy_fit1} (upper panel), the data and our best fit 
are displayed for the FUSE and \hst\ lines (listed also in Tab.~\ref{t:broad_lines}). 
The model seems to describe the broad line average emission reasonably well if $\gamma=-1.02\pm0.14$ 
(this is the 84\% confidence level for one interesting
parameter). The value of $\gamma$ is the only fitting parameter.
The integrated covering factor of the clouds set, 
evaluated only on the basis of the fitting of the hydrogen \Lya\ normalization, is 34$\pm$26\%.
Another parameter that was not fine-tuned in the fit is the outer radius of the BLR. 
The residuals to the best fit model are shown in Fig.~\ref{f:cloudy_fit1} (lower panel). 
Within $\sim 2\sigma$ the observed line luminosities agree with the LOC model.\\ 
Other complicating factors are likely to contribute to the observed BLR emission line spectrum. 
For example, the abundances may be different than solar, the line emitting clouds
 may have a wider range of column densities, 
or the geometry of the system maybe different than the simple spherical symmetry assumed here. 
However here we seek a simple parameterization that explains the bulk of the BLR emission. A 
detailed model of the UV lines produced by the BLR would be beyond the scope of this paper.\\ 
%
%
The luminosities of the major X-ray lines, 
as predicted by the BLR model are listed in Tab.~\ref{t:x_cloudy}. 
In this table, the luminosity of the Gaussian profiles 
with which the LETGS spectrum was fitted 
without using a synthetic
model are listed again, for comparison. 
We stress that these lines were not included in the LOC model, which was based instead on the UV data only. 
This
choice was made to minimize possible errors that a fit based on the X-ray data 
(complicated by the
continuum level and the absorption lines) would have brought to the LOC model fitting. 
The luminosities predicted
by the LOC model are consistent, within the errors, with a crude line by line fitting. 
In Fig.~\ref{f:prof_r_n} the best fit radial profile of the intrinsic line 
luminosity, integrated along the density ($n^{-1}$) is
displayed for eight interesting ions. 
The diamond points locate the luminosity-weighted mean radius \citep[e.g. ][]{bottorrf02}:\\

\begin{equation}
R_{\rm lw}\propto\frac{\int\int L(r,n)\ r^{\gamma+1}\ n^{\beta}\ dr\ dn}{\int\int L(r,n)\ r^{\gamma}\ n^{\beta}\ dr\ dn}. 
\end{equation}
 
In the figure, the relevant X-ray ions are included along with \ovi, which lies in the FUSE bandpass, and \civ, \nv\ 
and H\,\Lya, observed by HST-STIS. 
Most of the predicted luminosity should arise from clouds located at $r\sim 10^{16.3-17.5}$\,cm. 
The UV ions seem to be confined to radii $r>10^{17}$\,cm, while not unexpectedly, 
the mean emitting radius for the X-ray lines would lie 10 times further in. Extrapolating the model, 
we see that in principle the X-ray lines, in contrast to most UV ions, have a non negligible luminosity tail from 
radii smaller than 10$^{16}$\,cm.

\begin{table*}
\caption{\label{t:broad_lines} Rest wavelength, along with the observed 
luminosity and the luminosity predicted by the LOC model 
of the broad UV lines as measured by FUSE (labeled 1) and HST-STIS (labeled 2). For each observed 
blended-line, the value of Lum$_{\rm LOC}$ consistently refers to the summed contribution of the lines in the blend.}
\begin{center}
\begin{tabular}{lllllll}
\hline\hline
&  Ion  & Wavelength  & Lum$_{\rm obs}$ 		& Lum$_{\rm LOC}$ & Inst & Notes\\
&	& (\AA )	& ($10^{41}$erg s$^{-1}$)& ($10^{41}$erg s$^{-1}$)&	 &\\
\hline
&\svi\ & $933+944$	& $<$2.9	& 5.7				& 1 &	\\		
&\ciii\ & 977		& 10.7$\pm2.5$	& 68.0				& 1&	\\
&\niii\ & 989+991	& $<$2.52	& 2.2				& 1&	\\
&\Lyb\ + \ovi\    & $1025+1032+1038$	& 31$\pm6$	& 30.4		& 1&	\\
&\heii\	& 1085		& 1.6$\pm0.6$		& 2.4				& 1&	\\
&\Lya\	& 1216		& 75.7$\pm 19.4$	& 75.7				& 2& 1	\\
&\nv\	& $1238+1242$	& 13.6$\pm 6.0$		& 13.4				&2&	\\
&\oi\ +\siII\	& 1304	& 3.1$\pm 0.6$ 	&3.7				& 2&	\\
&\cii\		& 1335	& 1.3$\pm 0.4$	&2.0				&2&	\\
&\siiv\		& $1393+1403$& 11$\pm2$	&13.7				&2& 2	\\
&\civ\		&$1548+1551$&	70$\pm 12$ 	& 42.9				&2&	\\
&\heii\		&1640	&	17$\pm 4$	& 13.8			&2&	3\\

\hline

\end{tabular}
\end{center}

Notes: (1) Blended with \ov] (1218\,\AA) and \heii\ (1216\,\AA), (2) blended also 
with the \oiv] quintuplet (1402\,\AA) and the \siv] quintuplet (1406\,\AA), (3) 
blended with the \oiii] doublet (1661$+$1666\,\AA).\\


\caption{\label{t:x_cloudy} Rest wavelengths and 
intrinsic luminosity (in units of $10^{41}$ erg s$^{-1}$) of the lines 
in the LETGS (as reported in Tab.~\ref{t:gaussian}) compared with the X-ray line luminosities 
as predicted by the LOC model (last column) are listed. Note that the observed X-ray line luminosities were not used in the LOC fitting, 
which was based
on UV data only. Here the triplets location has been identified with the wavelength of the forbidden line.}
\begin{center}
\begin{tabular}{lllll}
\hline\hline
&  Ion  & Wavelength  & Lum$_{\rm obs}$ & Lum$_{\rm LOC}$		\\
&    & (\AA)  & (10$^{41}$ erg s$^{-1})$ & (10$^{41}$ erg s$^{-1})$		\\
\hline
&\cv$^1$ 	& 41.47		& $<3$	    	    & 0.3		\\
&\cvi\					&33.73		&$1.1\pm0.5$	    & 0.6		\\
&\nvi$^1$					&28.78		&$<1.1$	    	    & 0.2		\\
&\nvii\					&24.78		&$<1.7$	    	    & 0.3		\\
&\ovii$^1$	&$22.09$	&$5.0\pm0.8$	    & 4.8		\\
&\oviii$^2$   	&$18.96$	&$2.4\pm1.0$	    & 1.0		\\
&\neix$^1$	& 13.44		&$<2.5$	    	    & 0.8		\\
&\nex\ 					& $12.13$	&$\dots$    	    & 0.1		\\

\hline

\end{tabular}
\end{center}
$^1$ Triplets,\\
$^2$ blended with the \ovii\ He$\beta$.

\end{table*}


Based on these results, we included in the final X-ray spectral fit six Gaussian profiles with luminosity fixed at 
the value predicted by the LOC model.
The widths of the individual lines were set to the maximal value found in the UV (FWHM=9500\,\kms), very similar to the value 
used in the line-by-line fit (Sect.~\ref{par:gauss}). For the triplets, the blend of the three lines set 
the total FWHM. For \ovii, we used instead the same FWHM as determined in the phenomenological fit (Tab.~\ref{t:gaussian}).
The wavelengths were left free to adjust in a range of $\pm$0.2\AA\ around the centroid of the main line, mainly 
to take into account the line blending for triplets. The modeled line profiles are
shown in the lower panel of Fig.~\ref{f:best_fit}. 
The following analysis on the absorption spectrum was performed after the inclusion of the broad emission lines 
as predicted by the LOC model.

\subsubsection{Narrow emission lines and RRCs}\label{par:narrow}
Narrow emission features are not a dominant component of the total spectrum of \mrk. The flux 
of the \ovii\ forbidden line and the \oviii\ \Lya\ are $1.0\pm0.6\times10^{-14}$ and $7\pm5\times10^{-15}$
erg\,cm$^{-2}$\,s$^{-1}$, respectively. Finally, the flux of the \ovii\ He$\beta$ line is $4\pm3\times10^{-15}$
erg\,cm$^{-2}$\,s$^{-1}$.
Carbon recombination lines are not visible in the total spectrum. 
Radiative recombination continua (RRCs) of \cv\ and \cvi\ 
(at $\lambda_{\rm obs}$=32.57 and 26.05\,\AA, respectively, 
Fig.~\ref{f:best_fit}) are also seen. 
The temperature of the \cv\ RRC is $kT=1.7^{+1.0}_{-0.5}$\,eV, with an emission measure 
$EM=2.7\pm2\times10^{61}$ cm$^{-3}$. For the \cvi\ RRC, $EM<1.5\times10^{61}$ cm$^{-3}$, 
fixing the temperature at the \cv\ value.

\subsection{The absorbed spectrum}\label{par:abs}

The LETGS spectrum shows absorption lines, at the redshift of the source, over the $\sim5-45$~\AA\ band,
indicating that a gas with a range of ionization stages must be present.  
In Tab.~\ref{tab:lines} we list these lines, reporting the theoretical wavelength and the measured equivalent width
(EW).
The H-like and He-like ions of C, N and O are generally well determined. We report also the formal
equivalent widths for several important diagnostic lines that are not
significantly detected.  These measurements help constraining column
densities and the determination of the global properties of the absorber.
The inferred velocity dispersions are affected by large errors for most of the ions considered. 
Therefore we took into account only the strongest ions: \ovii , \ovi ,
\cvi\ and \cv\ to calculate a mean velocity dispersion $\sigma=46\pm21$ km s$^{-1}$, using a curve of growth
method. This value is in rough 
agreement with FUV values as measured by FUSE (G05). 
The velocity dispersion found for FUV absorption lines is 50 km s$^{-1}$. This is the value 
we adopt in the modeling.\\
For the lines whose EW is better determined, we fitted also the blue-shift with respect to the laboratory wavelength
(Tab.~\ref{tab:lines}). 
The uncertainty on the blue-shift is proportional to $\Delta v/({\rm S/N})$ 
where $\Delta v$ is
the velocity resolution at a given wavelength and S/N is the signal-to-noise ratio of the detected feature. 
For the lines located at longer wavelength, the shift measurement is in principle more precise. 
However, the higher resolution in these regions is compensated by a lower S/N ratio. 
Deriving  the covering factor of the ionized absorbing gas from the X-ray data is a difficult task, due to the
insufficient energy resolution. The most simple case is to assume a covering factor of one.  
The result of this fit is shown in Tab.~\ref{tab:lines} (model M1) where the EW and the column density of the most prominent lines are
listed. The fitting was performed using the SLAB model in SPEX, which 
calculates the transmission through a thin layer of gas, making 
no assumption on the underlying ionization balance \citep{kaastra02}. In this model the tunable parameters are: the
individual ionic column densities, the width of the lines, the outflow velocity and the covering factor.  
However, for this observation of \mrk, 
additional information is provided by the simultaneous UV data. 
In particular, the UV absorber for \mrk\ shows evidence of two velocity components (indistinguishable with the X-ray
resolution) whose covering factor differs from
unity by 10\% and 7\% respectively. Here we test this model using the LETGS data. 
The model M2 in Tab.~\ref{tab:lines} 
assumes two components (i.e. two SLAB components), as found from the FUSE analysis (G05). 
In the first component, the velocity dispersion is 
set to $\sigma$=50 km s$^{-1}$ with an assumed covering factor of 0.9. The second 
component contains 4\% of the column density of the first component, 
the velocity dispersion is 28 km s$^{-1}$ and the covering factor is 0.93. 
The predicted EW and the derived column densities for the two models 
are compared in the table. With the present LETGS data we are not able to distinguish between the 
two approaches (M1 and M2) and the X-ray absorber is consistent
to cover all the line of sight. This is reasonable, as the X-ray
source may be smaller than the UV source such that the ionized gas sees it as point-like. 
For simplicity we will assume that the
X-ray gas has a covering factor of one.\\
In the following, we describe two viable models to physically describe the ionized absorption in \mrk: separate gas
components differing in ionization parameter, column density and outflow velocity (Sect.~\ref{par:wa}) and a continuous
column density distribution as a function of the ionization parameter (Sect.~\ref{par:warm}). 
\begin{table*}
\caption{\label{tab:lines}
Laboratory wavelength, formal EW and outflow velocity for the X-ray absorption lines.
The EW and the column densities predicted by models M1 and M2 assume different covering factors 
for the absorbing gas and are discussed in the text. }
\begin{center}
\footnotesize
\begin{tabular}{llllllllll}
\hline
\hline
Ion & transition & $\lambda$ & EW     	& v$_{\rm out}$ & EW        &	  Log N$_{\rm ion}$   & EW   &	  Log N$_{\rm ion}$   & Notes\\ 			     
    &            & (\AA)     & (m\AA) 	& km s$^{-1}$ & (m\AA)    &	  ${\rm cm}^{-2}$   & (m\AA) & 	  ${\rm cm}^{-2}$&\\				     
    &            &           & obs.   	& obs. & M 1       &	  M 1	   & M 2 &	  M 2   &	   \\					     
\hline																		    
\ion{C}{iv}    &                         & 41.42     &  6$\pm$17 &		    & 8  &   15.13$\pm$1.14	   &  8    &	 15.16$\pm$1.11     		 & 1 \\  
               &                         & 40.94     &  43$\pm$29 	 &		    & 1 &		   &  1    &		   	    		& 1					       \\   
\ion{C}{v}     & 1s$^2$ - 1s2p $^1$P$_1$ & 40.268    & 29$\pm$20 & --380$\pm70$	    & 40 &	    16.92$\pm$0.34 & 48    &	 16.88$\pm$0.31     			  & 1 \\ 
               & 1s$^2$ - 1s3p $^1$P$_1$ & 34.973    &  25$\pm$8&			  &  27 &		   & 27    &					& 2					       \\ 
               & 1s$^2$ - 1s4p $^1$P$_1$ & 33.426    &  14$\pm$7 &			  &  18 &		   & 19    &					&				 \\	 
               & 1s$^2$ - 1s5p $^1$P$_1$ & 32.754    &  21$\pm$7 &			  &  12 &		   & 13    &					&				 \\	 
               & 1s$^2$ - 1s6p $^1$P$_1$ & 32.400    &   7$\pm$7 &			  &   8 &		   &  9    &					&				 \\	   
\ion{C}{vi}    & 1s - 2p (Ly$\alpha$)    & 33.736    & 36$\pm$8  & --390$\pm$140    & 33   &	    17.1$\pm$0.35  & 37    &	 17.02$\pm$0.30     			  & 2 \\ 
               & 1s - 3p (Ly$\beta$)     & 28.466    &   12$\pm$6	&		   & 19 &		   & 20    &					&				 \\	 
               & 1s - 4p (Ly$\gamma$)    & 26.990    &   20$\pm$6	&		   & 11 &		   & 13    &					&				 \\	 
               & 1s - 5p (Ly$\delta$)    & 26.357    &    7$\pm$6	&		   &  7 &		   &  8    &					&				 \\	 
               & 1s - 6p (Ly$\epsilon$)  & 26.026    &   -4$\pm$11	&		   &  4 &		   &  5    &					&				 \\	   
\ion{N}{v}     &                         & 29.42     &  4$\pm$8  &		    &  4  &	    $<$15.9	   &  4    &	 $<$15.9   	    		  &   \\   
\ion{N}{vi}    & 1s$^2$ - 1s2p $^1$P$_1$ & 28.787    & 28$\pm$6  & --320$\pm$180    & 22   &	    16.30$\pm$0.36 & 23    &	 16.34$\pm$0.34     			  &   \\ 
               & 1s$^2$ - 1s3p $^1$P$_1$ & 24.900    &   13$\pm$6      &		  &  10 &		   & 10    &					&				 \\	 
               & 1s$^2$ - 1s4p $^1$P$_1$ & 23.771    &   -8$\pm$6      &		  &   5 &		   &  5    &					&				 \\	   
\ion{N}{vii}   & 1s - 2p (Ly$\alpha$)    & 24.781    & 15$\pm$6  & --320$\pm$130    & 14  &	    16.16$\pm$0.63 & 14    &	 16.18$\pm$0.57     			  &   \\   
               & 1s - 3p (Ly$\beta$)     & 20.910    &  	 &		    &	      & 		   &	   &		            		& 3 \\   
               & 1s - 4p (Ly$\gamma$)    & 19.826    & -2$\pm$6  &		    &  1  &			   &  1    &		            		&   \\   
\ion{O}{i}     & 1s - 2p                 & 23.52     &  4$\pm$6  &		    &  4  &	    $<$16.5	   &  4    &	 $<$16.6   	    		  &   \\   
\ion{O}{ii}    & 1s - 2p                 & 23.30     & -1$\pm$6  &		    &  0  &  $<$16		   &  0    &	 $<$16              		&   \\   
\ion{O}{iii}   & 1s - 2p                 & 23.11     &  4$\pm$6  &		    &  4  &  $<$16.5		   &  4    &	 $<$16.5            		  &   \\   
\ion{O}{iv}    & 1s - 2p                 & 22.69     & 15$\pm$7  &		    & 15   &	    16.51$\pm$0.77 & 15    &	 16.53$\pm$0.68     			  & 4 \\   
\ion{O}{v}     & 1s - 2p                 & 22.374    & 18$\pm$6  &		    & 13   &	    16.10$\pm$0.51 & 13    &	 16.14$\pm$0.50     			  &  \\   
               & 1s - 3p                 & 19.92     & -9$\pm$7  &		    &  4   &			   &  4    &		            	  &   \\   
               & 1s - 4p                 & 19.33     &  6$\pm$6  &		    &  2  &			   &  2    &		            	  &   \\   
\ion{O}{vi}    & 1s$^2$2s - 1s2s($^1$S)2p& 22.01     & 19$\pm$8  &$+$28$\pm$120	    & 19   &	    16.72$\pm$0.45 & 20    &	 16.72$\pm$0.43     			  &  \\   
               & 1s$^2$2s - 1s2s($^3$S)2p& 21.79     &  2$\pm$7  & 		    &  9    &			   &  8    &		            	  &  \\   
               & 1s$^2$2s - 1s2s($^3$S)3p& 19.34     & 16$\pm$6  &		    &  9   &			   &  9    &		            	  &   \\   
\ion{O}{vii}   & 1s$^2$ - 1s2p $^1$P$_1$ & 21.602    & 28$\pm$7  & --120$\pm$80	    & 21   & 17.0$\pm$0.4	   & 25    &	 17.07$\pm$0.35     		  &  \\   
               & 1s$^2$ - 1s3p $^1$P$_1$ & 18.629    & 17$\pm$5  &		    & 13  &			   & 14    &		            	  &   \\   
               & 1s$^2$ - 1s4p $^1$P$_1$ & 17.768    & 10$\pm$5  &		    &  9  &			   &  9    &		            	  & \\   
               & 1s$^2$ - 1s5p $^1$P$_1$ & 17.396    & -1$\pm$5  &		    &  6   &			   &  6    &		            	  &   \\   
\ion{O}{viii}  & 1s - 2p (Ly$\alpha$)    & 18.969    & 17$\pm$5  & --440$\pm$170    & 13   & 16.45$\pm$0.48	   & 13    &	 16.50$\pm$0.47     		      &   \\   
               & 1s - 3p (Ly$\beta$)     & 16.006    &  1$\pm$5  &		    &  4   &			   &  4    &		              &   \\   
               & 1s - 4p (Ly$\gamma$)    & 15.176    & -4$\pm$5  &		    &  2   &			   &  2    &		              &   \\   
\hline
\end{tabular}
\end{center}
\end{table*}
\begin{table*}
\addtocounter{table}{-1}
\caption{Continued.}
\begin{center}
\begin{tabular}{llllllllll}
\hline\hline
\ion{Ne}{vi}   & 1s - 2p                 & 14.02     &  3$\pm$5  &		 &  3	& $<$16.7		 &  3	&     $<$16.7	 	        	  &   \\   
\ion{Ne}{vii}  & 1s - 2p                 & 13.81     &  7$\pm$5  &		 &  7	& $<$16.8		 &  7	&     $<$16.8	                	  &   \\   
\ion{Ne}{viii} & 1s$^2$2s - 1s2s($^1$S)2p& 13.65     &  7$\pm$5  &		 &  6	& 16.1$\pm$0.9  	 &  6	&     16.1$\pm$0.87	        	  &   \\   
               & 2s - 4p                 & 67.382    & 21$\pm$25 &		 & 25	&			 & 25	&			          &   \\   
\ion{Ne}{ix}   & 1s$^2$ - 1s2p $^1$P$_1$ & 13.447    & 10$\pm$4  &--270$\pm$500	 &  8	& 16.23$\pm$0.65	 &  8	&     16.27$\pm$0.61	        	  &   \\   
               & 1s$^2$ - 1s3p $^1$P$_1$ & 11.547    & -2$\pm$5  &		 &  2	&			 &  2	&		                  &   \\   
\ion{Ne}{x}    & 1s - 2p (Ly$\alpha$)    & 12.134    & -2$\pm$5  &		 &  0	& $<$16.23		 &  0	&     $<$16.25  	        	  &   \\   
\ion{Mg}{x}    & 2s - 3p $^2$P$_{1/2}$   & 57.920    & 11$\pm$13 &		 & 14	& 15.75$\pm$0.49	 & 14	&     15.76$\pm$0.48	        	  &   \\   
               & 2s - 3p $^2$P$_{3/2}$   & 57.876    & 24$\pm$12 &		 & 22	&			 & 22	&		      	          &   \\   
\ion{Si}{viii} & 2p - 3d $^4$P$_{5/2}$   & 61.070    & 18$\pm$12 &		 & 18	& $<$15.6		 & 18	&     $<$15.6	                	  & 5 \\   
\ion{Si}{ix}   & 2p - 3d $^3$D$_1$       & 55.305    & 25$\pm$10 &--600$\pm$300	 & 33	&	 15.23$\pm$0.52  & 30	&     15.26$\pm$0.48	        		  &  \\   
\ion{Si}{x}    & 2p - 3d                 & 50.524    & -1$\pm$12 &		 &  0	&	 $<$15.0	 &  0	&     $<$15.1	                	  &   \\   
               & 2s - 3p $^2$D$_{3/2}$   & 47.489    & -3$\pm$7  &		 &  0	&			 &  0	&		                  &   \\   
\ion{Si}{xi}   & 2s - 3p $^1$P$_1$       & 43.762    &  5$\pm$6  &		 &  5	&	 $<$15.4	 &  5	&     $<$15.4	                	  &   \\   
\ion{Si}{xii}  & 2s - 3p $^2$P$_{1/2}$   & 40.911    & 16$\pm$23 &		 & 16	&	 $<$16.51	 & 16	&     $<$16.53  	        	  & 1 \\   
               & 2s - 4p $^2$P$_{1/2}$   & 31.015    &  5$\pm$7  &		 &  5	&			 &  5	&		      	          &   \\
\hline\noalign{\smallskip}
\end{tabular}
\end{center}
\begin{list}{}{}
\item[1] In instrumental \ion{C}{i} edge
\item[2] Slightly broadened
\item[3] Blended with the Galactic \ion{O}{vii} resonance line
\item[4] Wavelength uncertain
\item[5] fit included other triplet lines with predicted ratio
\end{list}
\end{table*}
%
%
%
\subsubsection{A two ionization components model}\label{par:wa}

\begin{figure*}
\begin{center}
{\includegraphics[angle=90, height=5.8cm,width=15cm]{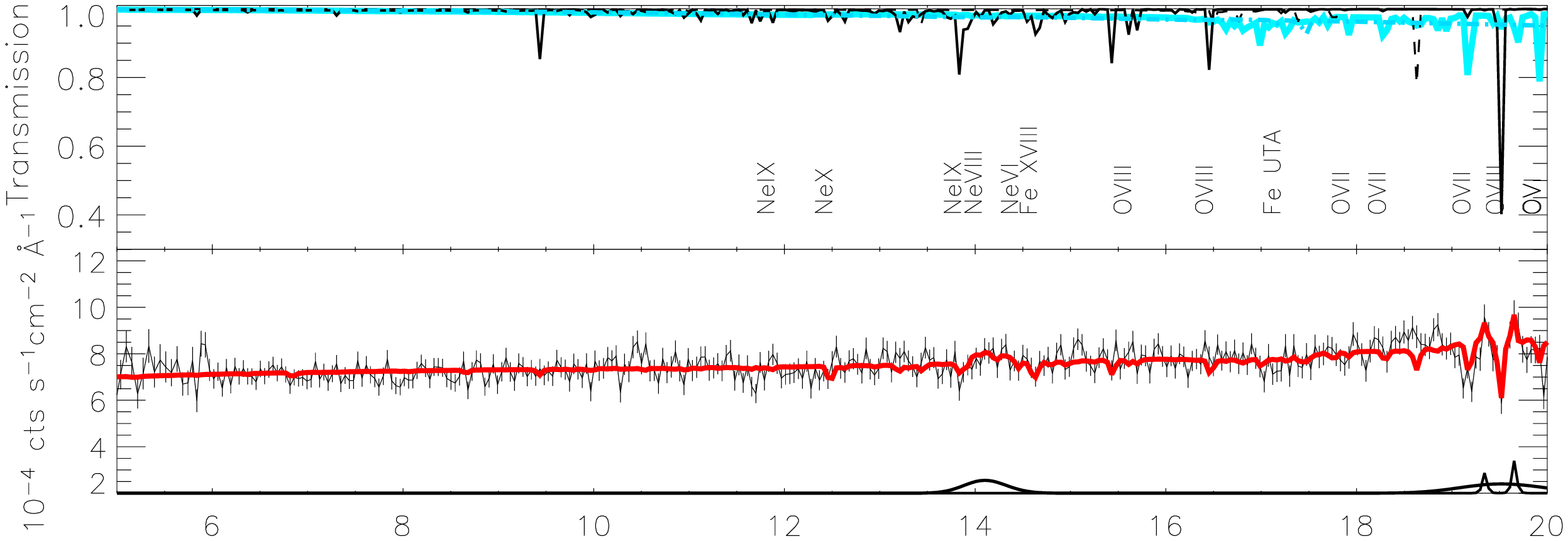}}
{\includegraphics[angle=90, height=5.8cm,width=15cm]{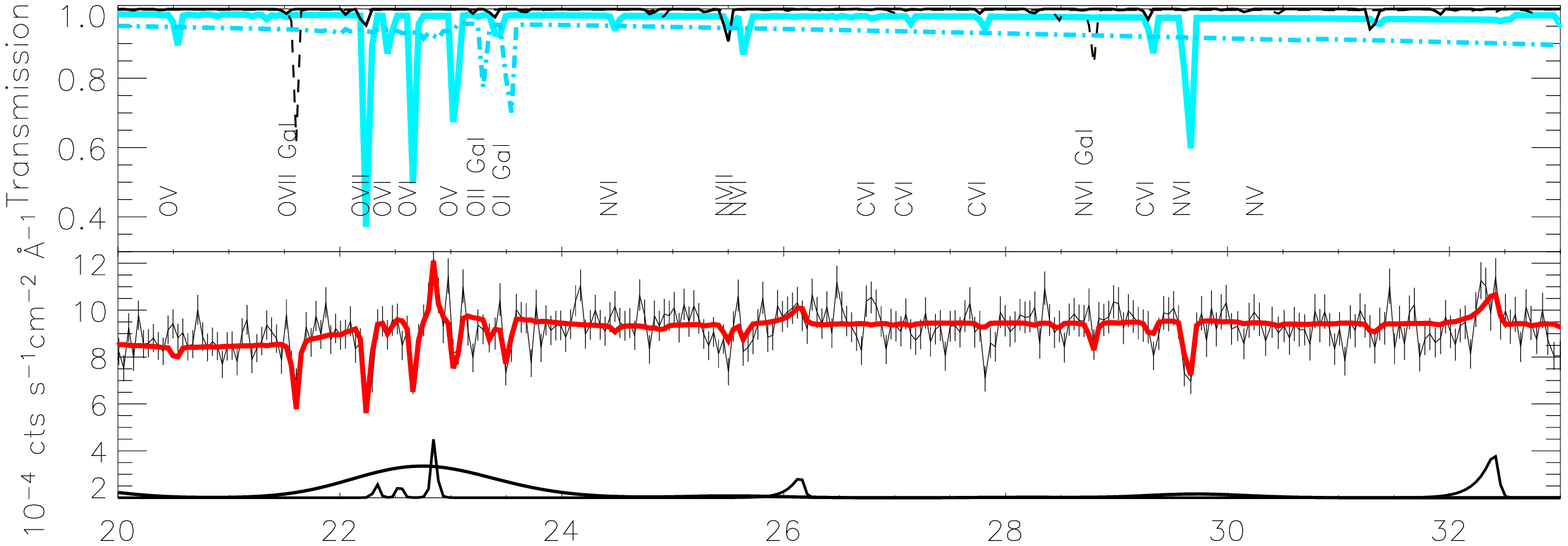}}
{\includegraphics[angle=90, height=5.8cm,width=15cm]{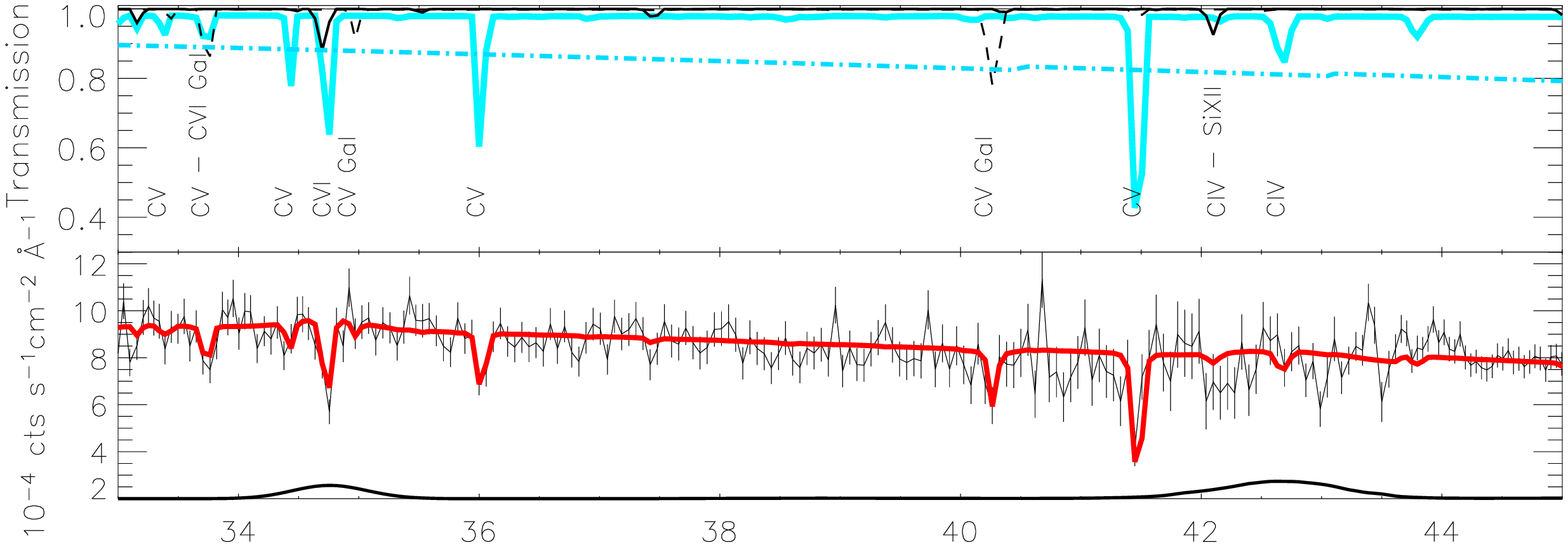}}
{\includegraphics[angle=90, height=5.8cm,width=15cm]{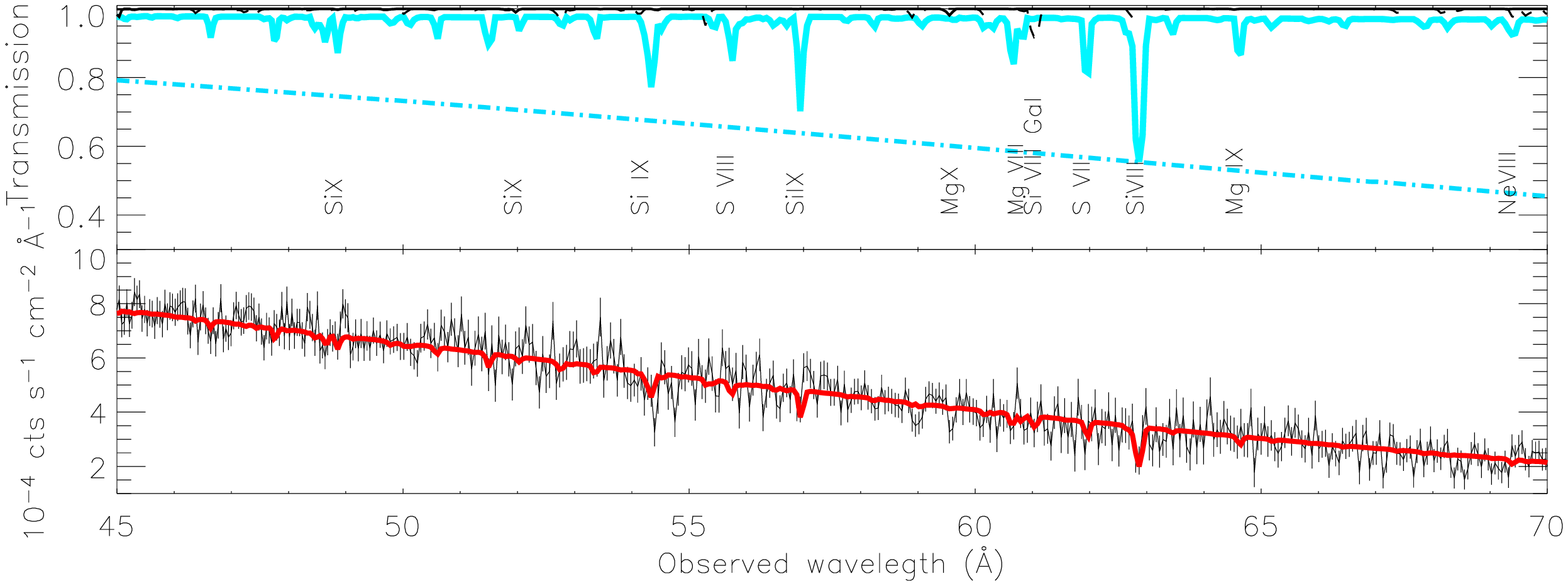}}
\end{center}
\caption{\label{f:best_fit} Lower panel: best fit to the LETGS spectrum of \mrk\ divided per wavelength range. 
The broad emission, predicted by the LOC fitting, and the narrow emission features, 
shifted up of a factor 2 for plotting purpose, are also overplotted. 
Upper panel: absorbing components: Dark dashed and light dot-dashed line: z=0 absorbers. Warm absorber: 
 light solid line ($\log\xi=0.47$), dark solid line ($\log\xi=2.49$). }
\end{figure*}

\begin{figure}
\begin{center}
\resizebox{\hsize}{!}{\includegraphics[angle=90]{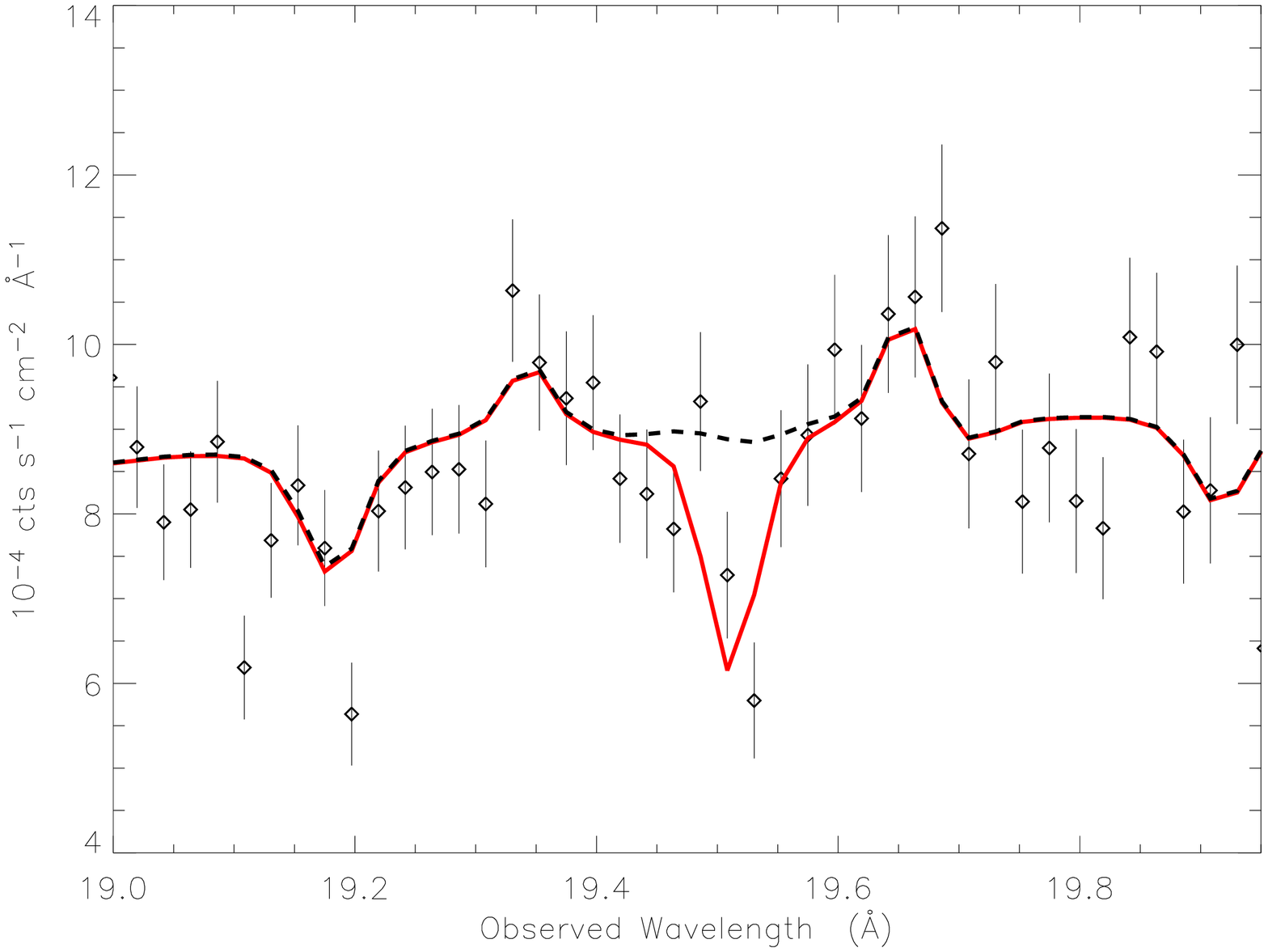}}
\end{center}
\caption{\label{f:oviii_one} Detail of the \oviii\ absorption line, fitted with only one component ($\log\xi=0.47$,
 dashed line) and with two components ($\log\xi=0.47,2.49$, solid line).}
\end{figure}

We modeled the data in terms of physical warm absorber components. 
For this purpose, we used the XABS model in SPEX. 
From an input spectral energy distribution (Fig.~\ref{f:sed}) for 
the ionizing continuum of the source, 
the model interpolates over a large grid of values for $\xi$ and $N_{\rm H}$, pre-calculated using Cloudy.  
Abundances are assumed to be solar \citep{grevesse98}. 
The
data require at least two ionized absorbers located at the redshift of the source. 
In Tab.~\ref{t:abs} we list the hydrogen column density, the ionization parameters and the outflow velocity 
of the absorbing gas. 
The best fit obtained with this model is displayed in the lower panel of Fig.~\ref{f:best_fit}. 
In the upper panel the transmitted spectrum is displayed. 
The light solid line refers to the lower ionization absorber while the higher ionization
gas is highlighted with a dark solid line.
The low ionization absorber (component 1 in Tab.~\ref{t:abs}) is characterized by a 
ionization parameter $\log\xi\sim0.47$ and it is 
tightly determined by the fit of the strong absorption lines of \cv, \cvi, 
\nvi, \ovi\ and \ovii\ (Fig.~\ref{f:best_fit}).    
\begin{table}
\normalsize
\caption{\label{t:abs} The two absorption components at the redshift of \mrk. For each
component we list the value of the ionization parameter 
$\log\xi$, the column density $N_{\rm H}$ and the outflow velocity $v_{\rm out}$.}
\begin{center}
\begin{tabular}{llll}
\hline\hline
&    $\log\xi$     &		$N_{\rm H}$	& $v_{\rm out}$\\
&   erg s cm$^{-1}$&	    $10^{20}{\rm cm}^{-2}$ & km s$^{-1}$\\	 
\hline
1 & $0.47\pm0.07$      & $1.23\pm0.23$         &$-202\pm50$\\
2 & $2.49\pm0.07$	       &$3.2\pm0.8$	       &$-560\pm130$    \\
\hline
\end{tabular}
\end{center}
\end{table}
The other component (labeled 2 in Tab.~\ref{t:abs}) is slightly faster and more ionized ($\log\xi\sim2.49$). It is 
mainly marked by the \oviii\ line, but also other high ionization absorption lines, detected with a lower significance, 
like 
\fexvii-\fexix, \neix, \nex\, \mgxi\ and \sixiii, help in defining the parameters of this second component. 
Adding a second component improves the fit by $\Delta\chi^2/\Delta\nu=40/2$, corresponding to a significance $>99.5$\%.
Many of the lines of the high ionization 
system lie in a region with a lower velocity resolution ($\Delta v\sim 750$ km\,s$^{-1}$ at 20 \AA) 
and the determination of the
physical parameters is therefore affected by higher uncertainties. 
However, if only the $\xi=0.47$ absorber is considered, the \oviii\ \Lya\ line is poorly fitted (Fig.~\ref{f:oviii_one}). 
Finally, as seen in Tab.~\ref{tab:lines}, \oviii\ and \ov\ have 
comparable column densities (the ratio is $\sim$2) and
this cannot be reached with a  single ionization parameter. 
The column density of the two ionization components is
quite low: 1.23 and 3.2$\times10^{20}$\,cm$^{-2}$ for $\log\xi=0.47$ and $\log\xi=2.49$, respectively.


\subsubsection{Continuous distribution model}\label{par:warm}
\begin{figure}
\resizebox{\hsize}{!}{\includegraphics[angle=90]{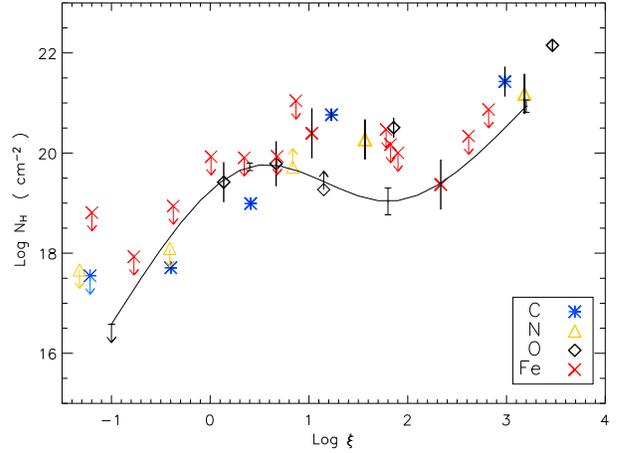}}
\caption{\label{f:ion_plot} The hydrogen column density as a function of the 
ionization parameter determined for: single ions (individual points) and an $N_{\rm H}$ 
continuous distribution model (solid line). See Sect.~\ref{par:warm} for a full description.}
\end{figure}

Alternatively to a discrete-components fit, 
the absorbed spectra can be modeled also as a continuous 
distribution of the hydrogen column densities of the ionized medium 
as a function of the ionization parameter. This is achieved using the WARM model in SPEX. 
The spectrum is fitted by a series of XABS models at intervals of 0.2 in $\log\xi$. 
At desired values of $\log\xi$, 
the error on the corresponding hydrogen column density ($N_{\rm H}^{\rm warm}$) is evaluated. 
We chose a $\log\xi$ interval 
between -1 and 3.2 and we evaluated $N_{\rm H}^{\rm warm}$ at those 2 points plus 2 points, equally spaced, in between.  
These four points (at $\log\xi=-1, 0.4, 1.8\ {\rm and}\ 3.2$, respectively), smoothly connected by the finer 
grid of $N_{\rm H}$, determined by the XABS series, are plotted in Fig.\ref{f:ion_plot} 
as a solid thick line.
Superimposed to that, we plot the derived
hydrogen column density for the more abundant elements: carbon, oxygen, nitrogen and iron. These column densities  
are again inferred using the SLAB model. 
The equivalent hydrogen column densities for each ion are then derived using solar abundances. 
For this fit we left out the
energies above 6 keV. In that region many K-shell transitions of iron 
ions are present, but due to the low effective area and possible calibration uncertainty the column
density of important ions as \fexxiii-\fexxiv\ would have been wrongly evaluated.
Moreover, low ionization ions of carbon (\ci-\civ) and nitrogen (\ni-\nv) are not well determined, as their
only feature in the X-ray band is the K edge. At the typical column density of the warm absorber of \mrk, the edge optical
depths of those ions is of the order of $10^{-(3-4)}$ only. Therefore, we chose not to include these ions in the SLAB fit. 
For the column density of \cii, \ciii, \nii\ and \niii, we used the values reported in \citet{scott04}. 
The ionic column density of \civ, \nv\ as well as \ovi\ could be taken from the simultaneous UV data (G05). 
In Fig.~\ref{f:ion_plot}, the hydrogen column density derived from 
the ionic concentrations for each ion is plotted against the $\log\xi$ values at which the ion is most likely formed. In
particular, the $\log\xi$ value for each specific ion $i$ is the result of an integration over large grid of $\log\xi$ values ($-8.5-6.5$) $vs$
ionic column density \citep[][]{steen05}. 
For many ions the column density cannot be determined with accuracy and there is a large scatter in the values. 
At $\log\xi$ in the range $0-1$, the WARM synthetic model is well traced by the single ions of \ov-\ovi. While the line for 
$\log\xi$ between 2.5 and 3 is represented mainly by the iron ions (\fexix-\fexx) and by the H-like ions of N and C. 
Carbon ions suffer from a high uncertainty, as some of
them lie close to the deep \ci\ instrumental edge. The two approaches that we described above do not deliver the same physical
picture. This will be discussed in Sect.~\ref{par:warm_disc}.

\subsubsection{Short-term variability in the warm absorber}
\begin{figure}

\begin{center}
\resizebox{\hsize}{!}{\includegraphics[angle=90]{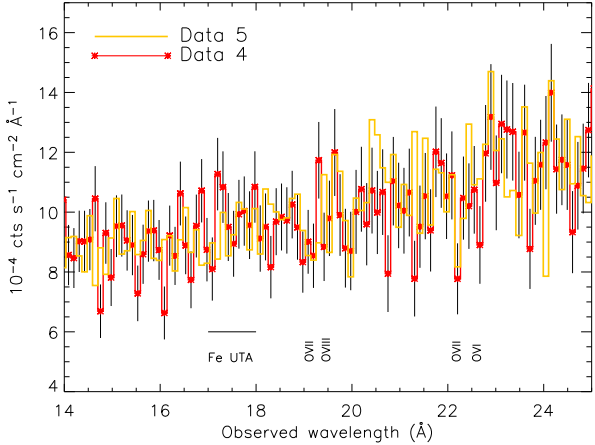}}

\end{center}
\caption{\label{f:wa_res} Comparison between the 4th and the 5th segment of \mrk\ observation
(Fig.~\ref{fig:lc}). The 4th spectrum (continuum line with asterisks) is normalized to the unabsorbed 
continuum of
the 5th observation. The 5th  spectrum is plotted with the light-solid line.}
\end{figure}

So far we have studied the combined spectrum of \mrk. A short term 
variability in the warm absorber in response to the modest continuum changes 
can in principle be detected. However, the analysis of
the imprint of the warm absorber, separately for the seven data sets, did not provide evidence of a statistically 
significant change. In Fig.~\ref{f:wa_res} we show, as an example, the comparison between the spectra taken during 
the 4th and 5th time intervals,
when the source underwent a major flux change, on a time scale of 2-3 days (Fig.~\ref{fig:lc}). The 4th
spectrum has been normalized by the higher flux unabsorbed continuum of the 5th observation. 
In this way any spectral modulation caused by a continuum change in flux and/or shape is canceled out and only the
information on the absorbers remain. 
In the continuum also
broad lines are included, as they are essential components in the evaluation of narrow absorption features. We see that
the oxygen complex does not change significantly. The iron UTA region 
($\sim 15-18$\,\AA, observed wavelength) is a useful tool to detect variations of the warm absorber
\citep[e.g, ][]{behar01,behar03,k03} as a small variation of the ionization parameter of the gas would shift the
iron UTA on the wavelength axis. No significant shift is detected in the data. 

\subsubsection{Absorption at redshift zero}\label{par:mw}
The spectrum of \mrk\ is also marked by narrow absorption lines which are consistent with being produced at zero redshift. 
Weakly ionized absorption, likely to arise in the ISM of our Galaxy is highlighted by the
\oi\ feature at 23.04\,\AA\ and a weaker \oii\ absorption line (Fig.~\ref{f:best_fit}). 
Such an absorption is well parameterized by a collisionally ionized gas with a very
low temperature ($3.1\pm1.2$ eV) and a column density of 0.72$\pm0.08\times10^{20}{\rm cm}^{-2}$.
This source is already known to show an ionized absorber at redshift zero
 \citep[traced by \ovi\ in the UV band, e.g.][]{savage}. Also in the
X-ray band we detect ionized material traced by 
several absorption features from \ovii, \nvi, \cv\ and \cvi\ (Fig.~\ref{f:best_fit}).   
We modeled this system of lines with a collisionally ionized plasma model with a temperature 
$kT=7.2\pm1.7$\,eV and a column density $N_{\rm H}=3.6\pm0.3\times10^{19}\ {\rm cm}^{-2}$.\\ 
This absorbing gas may be located in the environment of
the Milky Way \citep[e.g., ][]{sembach,wang}, in the form of high velocity clouds moving and interacting with each other in the
Galactic halo \citep{collins05}. Another interpretation locates the absorbing gas on 1--3 Mpc scale, in the local group
\citep{nicastro02}.
A study of the \ovi\ absorption line in the \mrk\ UV spectrum is presented by \citet{fox04}.
 A detailed interpretation of the X-ray absorbing components on the line of sight of \mrk\ is 
 discussed e.g. in \citet{will05}.  

\section{Discussion}

\subsection{Emission from the broad line region }\label{par:broad_disc}
The broad emission features detected in the LETGS spectrum of \mrk\ can be modeled in terms of 
emission lines from the BLR. 
Using the LOC model \citep{baldwin95}, we found that the UV lines are modeled by emission 
from clouds whose density and radial 
distribution follow a power law. The radial distribution is found to decrease with a slope 
$\gamma=-1.02\pm0.14$, while the slope for the density distribution was fixed to $-1$ (Sect.~\ref{par:broad_data}). 
The integrated covering fraction of
the clouds set, calibrated on the hydrogen \Lya\ fitting, is loosely constrained, being 34$\pm$26\%.   
These results are in agreement with what found for the composite quasar spectra \citep{baldwin97}.
The LOC model was also applied to another bright Seyfert~1 galaxy, \ngc, \citep{korista00}. In that case, the 
density and radial distributions  slopes $\beta=-1$, $\gamma\sim -1.2$, respectively, well explained the HST-STIS data. 
In the present paper for the first time the LOC model has been extended to the X-ray band. 
The intrinsic luminosity of the X-ray lines have been calculated from this model 
and applied to the LETGS data fitting. An independent fit of the X-ray spectrum using Gaussian profiles 
leads to flux estimates which are
consistent, within the errors, with the ones predicted by the LOC (Tab.~\ref{t:x_cloudy}), strengthening the validity of this approach.
In the case of \mrk, the \ovii\ triplet complex is the most prominent feature (30\% above the continuum, 
Fig.~\ref{f:best_fit}), as this oxygen ion is steadily produced in a wide range of physical conditions. 
The other broad X-ray lines are weaker ($\sim$10\% above the continuum), but 
help in modifying the continuum and better constrain the warm absorber parameters. For these lines we cannot draw firm
conclusions, as their significance in the data is relatively low.\\  
It is feasible that there may  be a part of the BLR which has higher ionization and that produces lines only 
visible in the X-ray band. This may give rise to some additional flux in the lines, 
possibly variable in time and thus appearing only in some time
segments, but its level is within the noise.
 In principle, measuring the amount of the excess would allow us to quantify 
the physical parameters of a highly ionized skin of the BLR. 
However it is extremely challenging, first because of the relatively low statistical 
significance of any additional excess,
second because we modeled the BLR using an average source flux that 
possibly introduces some more scatter in the predicted values of the X-ray lines luminosity. 
Being able to study a part of the BLR which only emits in the X-ray band would be indeed 
very important in understanding the stratification of the BLR and its 
velocity field \citep[e.g., ][]{baldwin97,2001MNRAS.328..409G}.
Following further the BLR interpretation, 
we find that the luminosity--weighted radii (Sect.~\ref{par:broad_data}) map a
wide region that extends from 10 to 100\,ld from the central source, which produces lines visible
both in UV and X-ray band. This size is smaller ($\sim$67\,ld) when only the UV lines are considered. 
This estimate is larger than 
the BLR size obtained by reverberation mapping studies \citep[$<30$\,ld,][and references therein]{stirpe}. 
However, we note that 
the ionization conditions within such an extended region are sensitive to the long-term flux history of the
source. In particular there is evidence that more than forty days before the present multiwavelength campaign, 
the V-band flux of \mrk\ was up to a factor of 8 lower (depending on the host galaxy subtraction, 
Gaskell et al. 2006, in prep.).   
From that epoch, the V band flux, which should be, on long time scales, correlated also with the
high energy flux, gradually rose to reach the higher state caught by LETGS-FUSE-HST. As the size of the BLR 
is proportional to
the square root of the ionizing luminosity \citep{peterson93}, the size estimated here could be reduced by a factor as
large as two or three.\\ 
The luminosity-weighted radius of the higher ionization X-ray ions is located at radii up to ten times
smaller than for the UV ions. If the motion of the BLR clouds is purely keplerian, this would imply a velocity 
broadening up to a factor of three larger. The only broad excess for which we measured the FWHM is the \ovii\ triplet blend 
(FWHM=1.9$^{+0.7}_{-0.4}\times10^4$\,\kms, Tab.~\ref{t:gaussian}) which is consistent, within the errors, 
with such a large broadening. Unfortunately in this data set this possibility cannot be tested 
on non-blended, higher ionization lines (like for instance \cvi).
The extrapolation of the BLR model down to small radii shows that the emission of the X-ray BLR has a non-negligible tail that goes 
down to $\sim$0.8 ld from the source, corresponding nominally to $\sim$300 Schwarzschild
radii, given the BH mass of \mrk\ 
\citep[$\sim 3\times10^{7} M_{\odot}$, ][]{1999ApJ...526..579W}. 
At this specific distance we would 
not expect a significant relativistic broadening of the line profile, 
but X-ray emission from highly ionized gas that is not efficiently producing UV lines
would be possible (Fig.~\ref{f:prof_r_n}).\\ 
The detection of relativistically broadened line profiles at soft X-ray energy has been claimed for a number of sources
\citep[e.g. ][]{branduardi,kaastra02,ogle04}. Our LETGS data do not show 
significant evidence of an asymmetric profile, especially at the 
wavelength of \oviii, \nvii, \cvi. Moreover, these profiles would be blurred by the wide, 
non-relativistic lines produced in the BLR. 
Among the H-like lines that we are able to detect in the LETGS band, \oviii\ lies 
in a privileged region, where the effective area is higher and the spectrum is not contaminated by instrumental
features. However, to model a skewed, relativistically broadened, profile for \oviii\ 
is difficult because of the absorption features of the iron UTA at observed wavelength of $\sim 17$ \AA\ for the
warm absorbers intrinsic to \mrk, and at $\sim 19$\,\AA\ for the ionized absorbers in our Galaxy. The structure 
of the iron UTA may not be yet completely accounted for in the models and may
cause additional uncertainties.

\subsection{The structure of the warm absorber}\label{par:warm_disc}
\begin{figure}
\resizebox{\hsize}{!}{\includegraphics[angle=90]{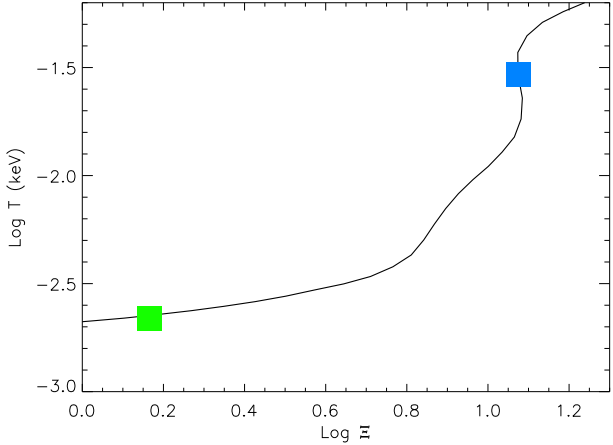}}
\caption{\label{f:lxi_t}The pressure ionization parameter {\it vs } the electron temperature. 
The two components of the warm absorber for \mrk\ are indicated as squares.}
\end{figure}

The spectrum of \mrk\ is absorbed by at least 
two absorption systems (Sect.~\ref{par:wa}). They show a significant blue-shift and a relatively 
high value of the ionization parameter and therefore can be 
unmistakably associated with the warm absorber seen in other 
Seyfert~1 galaxies (Tab.~\ref{t:abs}). 
Low ionization metals (e.g. \cii, \ciii, \nii) are marginally detected in the UV band, but for completeness, we 
included them in our analysis (Sect.~\ref{par:warm}). 
As discussed in \citet{scott04}, 
the association of these ions with the nuclear activity is unlikely. Because of their low ionization and the lack of any
blue-shift in the lines, they most probably arise in the \mrk\ host galaxy. 
The two high ionization components may be the only discrete constituents of the warm absorber 
\citep[e.g.][]{k05} or be only a part of a multi-ionization,
continuous outflow \citep{steen05}.  
A discrete-components scenario would be supported by finding that the observed 
components values have the same $\log\Xi$
in the curve of thermal stability, which is shown in Fig.~\ref{f:lxi_t}. 
Here we plot
 the log-log distribution of the pressure ionization parameter ($\Xi$) {\it vs} 
the electronic temperature ($T$) for \mrk. 
The value of $\Xi$ is defined as: $\Xi=L/4\pi r^2cp$, that is the ratio of photon pressure
to gas pressure. 
This may also be expressed as a function
of $\xi$ and the electronic temperature $T$: $\Xi=\xi/4\pi ckT$. The curve 
was obtained using Cloudy, calculating 
a grid of values of log$\xi$ and the relative electronic temperature of a thin 
layer of absorbing gas illuminated by an ionizing continuum. 
The shape of the curve is sensitive to the SED of the source (Fig.\ref{f:sed}).
The filled squares in Fig.~\ref{f:lxi_t} correspond to the log$\Xi$ values for the 
warm absorbers detected in \mrk. Given their position on
the S curve, these component cannot share the same pressure ionization parameter. 
This would have ensured a long-lived structure for a discrete-component warm absorber
unless other mechanisms, like magnetic confinement, are playing a role. 
If this were case, pressure
equilibrium would not be mandatory. 
Also a high amplitude variability of the warm absorber in response to a continuum flux variation 
could be in favor of a discrete-component
model \citep[e.g., ][]{k05}. 
Observationally, a small variation of the ionization parameter would cause a 
detectable shift in the iron M shell
UTA complex. In the case of \mrk\ this possibility cannot be tested, as the variation of the source 
is too small and the warm absorber too shallow to be able to detect a shift in the iron region (Fig.~\ref{f:wa_res}).

On the other hand, also a scenario that considers a 
continuous distribution over $\xi$ (Sect.~\ref{par:warm}) 
cannot be straightforwardly proven. In Fig.~\ref{f:ion_plot} we showed the
results of the WARM model, which mimics a continuous $N_{\rm H}$ distribution (Sect.~\ref{par:wa}). We see that 
the hydrogen column density distribution derived from single ions appears to give a different picture of a power law-like
distribution. Indeed, if we take into account
only the higher ionization ions ($\log\xi>0$), taking out all the upper limits from the fit, 
the data can be modeled by a power law with index $\alpha\sim 0.49$, where we consider $N_{\rm H}\propto\xi^{\alpha}$. 
This is very similar to what was found for \object{\ngc}, where \citet{steen05} estimated $\alpha\sim 0.40$ for this
range of ionization parameters ($\log\xi\sim-0.2-3.5$), despite that the column densities
for the \ngc\ warm absorber are a factor of $\sim 10$ larger. A similar trend is found for \object{NGC~4051}
\citep[$\alpha\sim 0.5$ ][]{ogle04}, with an apparent peak for $N_{\rm H}$ at $\log\xi=1.4$. We note however 
that this result is not completely comparable to \mrk\ and \ngc\ as  \citet{ogle04} compute $\log\xi$ at the peak of the
 ionization for each ion. Not taking into account that ions are formed in a range of $\log\xi$ can lead to
a quite different distribution. 
Finally, in the case of \mrk, the extrapolation of the $\log\xi>0$ power law to lower
ionization ions, excludes the lowest ionization ions 
from the continuous outflow structure, consistent with the idea that those ions are produced in a distant region.\\ 
An additional constraint is provided by the WARM model fit (Fig.~\ref{f:ion_plot}). The 
hydrogen column densities from the WARM model have more 
robust values, as the SPEX fit synthetically takes
into account all transitions for a given value of the ionization parameter. 
The WARM distribution does not suggest a straight power law fit. However, 
the power law fit to the single ions distribution ($\alpha\sim0.49$) nicely adapts to the
$\log\xi=$0.4 and 3.2 points at which the $N_{\rm H}$ was evaluated in the WARM model (Sect.~\ref{par:warm}). 
If we take this as the signature of a
continuous distribution, both the lower ionization end of this distribution ($\log\xi<-1$) and the point evaluated at 
$\log\xi=1.8$ deviate from the power law distribution (with 4$\sigma$ significance). In the framework of the structure of
the ionized outflow, we are mostly interested in the apparent dip in this continuous distribution at 
$\log\xi=1.8$. We further verified that fitting the data with an additional XABS components 
with $\log\xi$ constraint to lie between 0.47 and 2.49 (values from the XABS best fit, see Tab.~\ref{t:abs}) 
does not significantly change the goodness of fit in terms of $\chi^2$, but indeed provides a lower
 and significant value for 
$N_{\rm H}$. We find $\log\xi=1.5^{+0.3}_{-0.2}$ and $N_{\rm H}=2.8\pm0.4\times10^{19}$\,cm$^{-2}$.\\
The main point that may be deduced from this exercise is that 
the continuous distribution we tried to define is
in fact non-monotonous. 
We do not find a sharp bimodal distribution with $\log\xi\sim0.47$ and
$\log\xi\sim2.49$ being the only components of the warm absorber (as suggested by the XABS model, Sect.~\ref{par:wa}). 
Intermediate values may exist, but with a
column density that is low enough not to be easily detected. This is in contrast with 
what is predicted by a power law distribution fit.    
The statistics does not allow us to detail further these findings. For instance the kinematic
characteristics of this outflow are not precisely determined: the blue-shifts of the 
two ionization components we find are only marginally different (Tab.~\ref{t:abs}). 
The kinematic warm absorber structure for \mrk\ is indeed quite complex. A velocity-resolved spectroscopy 
of the UV absorption troughs of this source 
shows that not only partial covering plays a role \citep[][Sect.~\ref{par:abs}]{arav05}, 
but that abundances of C, N, and O can differ from solar values (Arav et al. 2006, in prep.). 
The N/O and C/O ratio delivered by the X-ray data are affected by large error bars, therefore a direct comparison with the UV
results is not
conclusive. 
  
\subsection{Density diagnostics}\label{par:ostar}
\citet{jelle04} discussed in detail the possibility of the detection of \ov\ absorption lines 
from a meta-stable level in \mrk. The main 
absorption feature (\ov$^{\star}$) lies at $\sim22.5$ \AA\ in the rest frame of the source (23.18 \AA\ in the observed
spectrum, Fig.~\ref{f:best_fit}).
This kind of transition of Be-like ions can take place only at particularly high densities. 
Therefore, a precise determination of the
physical parameters of such absorption features can serve as an important test for the distance determination of the 
warm absorber. Here we try to verify a connection between the parameters deduced from the \ov$^{\star}$ line and any
of the absorber components in \mrk.\\
The temperatures \citet{jelle04} infer for the gas producing the \ov$^{\star}$ line range between $2-4$ 
eV, corresponding, for the SED of \mrk, to $\log\xi\sim0-1.5$. 
This estimate of the ionization parameter, together with the hydrogen column density ($N_{\rm H}$) 
that we can derive from the equivalent width of the \ov\ ground state absorption lines 
 \citep[$N_{\rm OV}=1.4^{+0.5}_{-1}\times10^{16}\ {\rm cm}^{-2}$, ][]{jelle04}, 
 give already the basic physical quantities of a warm absorber. 
In order to ensure that the density value of a gas producing \ov$^{\star}$ is not unrealistic, 
we find that a gas temperature $kT<$3\,eV is needed. For this temperature, the \ov/O ratio is
0.02, implying an equivalent hydrogen column density $N_{\rm H}\sim8\times10^{20}\ {\rm cm}^{-2}$, 
which is still roughly compatible with the column density we find for a continuous-distribution warm absorber
model (Fig.~\ref{f:ion_plot}) . The associated 
ionization parameter would be $\log\xi\sim 1$. 
The comparison between the temperature of this gas ($\sim3$ eV) 
and the measured 
relative population of the 
\ov\ meta-stable
levels \citep[0.125--2,][]{jelle04}, provides an estimate for the density \citep[see Fig.~5, ][]{jelle04}. We obtain 
$n>3\times10^{12} {\rm cm}^{-3}$ and, as a consequence, 
a distance from the ionizing source  $r_0<3\times10^{15}\ {\rm cm}$, comparable with
the location of a more ionized part of the BLR. 
This estimate relies uniquely on the tentative
detection of \ov$^{\star}$. Further observational evidence is needed to support the identification of this line.      

\subsection{The emission counterpart of the warm absorber}
Whether the gas which is responsible for the blue-shifted absorption lines is also observed in emission, is a controversial
issue. The main reason is that the  absorption/emission connection is dependent on unknown parameters, such as the
density, the distance, and the overall geometry of the system. The emission counterpart of the X-ray 
warm absorber may be the
narrow lines produced in the NLR in a form of a bipolar cone with a wide opening 
angle \citep[e.g., NGC~3783, ][]{behar03}.
The observational facts that would support this scenario are: a lack of response of the warm absorber to the
central source variability and the similarity of the absorption and emission lines parameters, such as velocity width, column
density, ionization stage. In the case of \mrk, any suggestion of a 
short-term variation of the absorber parameters as a function 
of the ionizing flux is unfortunately too weak ($\sigma<2$) to be investigated quantitatively. 
Narrow emission lines are not
affecting the spectrum significantly. 
From the study of the possible \ov$^{\star}$ feature 
\citep[Sect.\ref{par:ostar}, ][]{jelle04}, we inferred an upper limit for the distance of the warm absorber 
($r_0<3\times10^{15} {\rm cm}$), at least for an intermediate ionization component. The emission 
counterpart of the X-ray warm absorber would be located at the same distance as the higher ionization BLR lines 
(i.e. the X-ray lines, Sect.~\ref{par:broad_disc}). Can the BLR cloud themselves produce the warm absorber? 
The particular blue-shift seen in the absorption feature can be line-of-sight dependent \citep{elvis}, while the broad lines can be emitted
over a maximal opening angle of $2\pi$. The UV broad lines are consistent to 
have a variety of velocity including extreme blue-shifted
components of $\sim$-2000 km s$^{-1}$ (G05). 
Moreover the absorbing and emitting gas can have the same structure, as 
the BLR ``clouds" can be organized in the form of a wind arising above the accretion disk
\citep[e.g., ][]{bottorff97} just like the warm absorber \citep[e.g., ][]{elvis}. 
However, one insurmountable discrepancy between the two media is the
measured column density. Indeed, the UV (and X-ray) broad emission lines can be produced for column densities down to 
$N_{\rm H}\sim10^{22}$\,cm$^{-2}$ \citep{kor97}, which is almost two orders of magnitude higher than the 
warm absorber of \mrk. Furthermore, the evidence of narrow absorption lines ($\sigma\sim50$\,km s$^{-1}$) point to a stable
flow. This would be difficult to maintain in a turbulent medium like the one producing the BLR. 
The
possibility that the broad emission lines are produced by the same gas component as the warm absorber is then ruled out. 
However, in a picture of a
windy gas above the accretion disk, our line sight may intercept first a portion of the absorber very close to the central source. Its
transverse thickness must be very small ($N_{\rm H}/n_{\rm H}\approx10^{5-6}$\,cm). 
This sheet of gas may be the outer, low 
column density part of the gas structure that produces the broad emission. 
If the limit on the distance of the absorber is
confirmed (Sect.~\ref{par:ostar}), this could be a way to make the 
two media coexist at approximately the same distance from the central source. A correspondence between the distance of the warm absorber and the BLR 
was also found in other X-ray sources \citep[e.g. NGC~3516, ][]{netzer02}. 
A physical connection between the BLR and the warm absorber 
was proposed for NGC~3783, observed by \ch-HETGS \citep{k03}. Finally, the gas temperature inferred from the RRCs fit translates in an ionization parameter, $\log\xi$, from --1.3 to
0.8, which is certainly consistent with the $\log\xi$ of one of the main absorbing components of the warm absorber in 
\mrk. Therefore, a physical link between the warm absorber and the gas producing the RRCs cannot be ruled out. 

\section{Conclusions}
We have presented the analysis and modeling of the data of \mrk, observed for 360\,ks by \ch-LETGS. 
For the first time, we have extended to the X-ray band the ``locally optimally cloud" model, first proposed by \citet{baldwin95} 
to describe the BLR emission
lines seen in the UV. This has been achieved by fitting first the luminosity of the UV lines, measured
simultaneously by FUSE and HST-STIS (G05) and deriving, from the best fit, the X-ray luminosity.\\
In agreement with the LOC model predictions, the distribution of the radial distance of the 
BLR ``clouds" follow a power law with index $\gamma=-1.02\pm0.14$, assuming a
density distribution with index $\beta=-1$. 
We find that the inferred X-ray line luminosity well describe the X-ray spectral shape. If, independently, we fit
the X-ray broad emission features with Gaussian profiles, the results are, within the errors, 
consistent with the UV modeling. The most evident X-ray broad emission feature, the \ovii\ triplet, which exceeds the
continuum of about 30\%, is well explained by the LOC model. 
The profiles of other important X-ray lines of the H-like and He-like C, N, O ions are less clear in the spectrum and
cannot be studied in more detail. 
It would be important to quantitatively measure the contribution of such ions, as they 
may be more sensitive to the physical conditions than \ovii, which is steadily produced for a wide range of gas densities and
distances from the ionizing source. 
There are not significant spectral residuals in excesses after the inclusion of the X-ray broad lines. 
This suggests first that the gas of the BLR emitting the bulk of the UV emission 
is also sufficient to explain the X-ray emission spectrum. Second, that the contribution of 
lines arising instead from the innermost region of the accretion disks, and thus relativistically broadened, is
negligible in the case of \mrk.\\ 
In the radial distribution of the line luminosity, the luminosity--weighted radius for each ion considered defines a wide 
region of roughly 90\,ld for the UV--X-ray BLR (67\,ld if only the UV ions are considered). 
This is larger than predicted by reverberation mapping 
studies \citep{stirpe}. However, our estimate could be lowered by a factor up to three noting that the BLR size 
is sensitive
to the long term variations of the continuum flux and that at the epoch 
of the present campaign the optical flux was at the higher end of a
gradual rise that lasted nearly two months.\\
A proper modeling of the broad emission lines in the LETGS spectrum helps in the analysis of the absorbed spectrum. 
We find the signature of at least two gas components which can be unmistakably associated with a warm absorber. 
However, there is some evidence that suggests that this may be only a partial description of the absorption.
First, absorption lines do not show any significant variation following the modest changes 
of the central source flux and, second, the two absorption components cannot be in pressure equilibrium.
This evidence, although not decisive, does not strongly support the hypothesis of a compact absorber \citep{k05}. 
We investigated the possibility that the absorbing gas is instead distributed over a wide range of ionization
parameters \citep{steen05}. We find that neither of the two models perfectly matches our data. A smooth
distribution of the column densities is accepted, but $N_{\rm H}$ does not monotonically increase 
as a function of $\xi$ (following for example a power law), 
but it is rather significantly bent.\\
From the exploratory study of the possible \ov\ absorption line to meta-stable level in the \mrk\ X-ray spectrum, \cite{jelle04} 
infer a density for the absorbing gas of the
order of $10^{14}$\,cm$^{-3}$. In the context of our analysis, we 
find that an intermediate ionization gas component ($\log\xi\sim 1$), with a column density roughly consistent with our
continuous-distribution model, could be consistent to produce the \ov$^{\star}$ line. This would imply 
an upper limit to the distance from the central source of $3\times 10^{15}$\,cm for the gas. 
Although a distance close to the BL emitting region for the warm
absorber is not new in the literature \citep[e.g., ][]{nicastro99,netzer02}, in the case of \mrk, a global model explaining
the coexistence of the complex warm absorber structure and the BLR cannot be easily depicted.       
\begin{acknowledgements}
The Space Research Organization of the Netherlands is supported
financially by NWO, the Netherlands Organization for Scientific
Research. 
\end{acknowledgements}


\begin{thebibliography}{}
\bibitem[Arav et al.(2005)]{arav05} Arav, N., Kaastra, J., 
Kriss, G.~A., Korista, K.~T., Gabel, J., \& Proga, D.\ 2005, \apj, 620, 665 
\bibitem[Bachev \& Strigachev(2004)]{bachev04} Bachev, R., \& 
Strigachev, A.\ 2004, Astronomische Nachrichten, 325, 317 
\bibitem[Baldwin et al. (1995)]{baldwin95} Baldwin, J., Ferland, 
G., Korista, K., \& Verner, D.\ 1995, \apjl, 455, L119 
\bibitem[Baldwin(1997)]{baldwin97} Baldwin, J.~A.\ 1997, ASP 
Conf.~Ser.~113: IAU Colloq.~159: Emission Lines in Active Galaxies: New 
Methods and Techniques, 113, 80 
\bibitem[Behar et al.(2001)]{behar01} Behar, E., Sako, M., \& 
Kahn, S.~M.\ 2001, \apj, 563, 497 
\bibitem[Behar et al. (2003)]{behar03} Behar, E. et al.\ 2003, \apj, 598, 232 
\bibitem[Blustin et al.(2005)]{blustin} Blustin, A.~J., Page, 
M.~J., Fuerst, S.~V., Branduardi-Raymont, G., \& Ashton, C.~E.\ 2005, \aap, 
431, 111
\bibitem[Bottorff et al.(1997)]{bottorff97} Bottorff, M., Korista, 
K.~T., Shlosman, I., \& Blandford, R.~D.\ 1997, \apj, 479, 200 
\bibitem[Bottorff et al.(2002)]{bottorrf02} Bottorff, M.~C., 
Baldwin, J.~A., Ferland, G.~J., Ferguson, J.~W., \& Korista, K.~T.\ 2002, 
\apj, 581, 932 
\bibitem[Branduardi-Raymont et al. (2001)]{branduardi} Branduardi-Raymont, G., Sako, M., Kahn, S.M., et al., 2001, A\&A 365, L140
\bibitem[Brinkman et al. (2000)]{2000ApJ...530L.111B} Brinkman, A.~C., et 
al.\ 2000, \apjl, 530, L111 
\bibitem[Collins et al.(2005)]{collins05} Collins, J.~A., Shull, 
J.~M., \& Giroux, M.~L.\ 2005, \apj, 623, 196 
\bibitem[Crenshaw et al.(1999)]{crenshaw99} Crenshaw, D.~M., 
Kraemer, S.~B., Boggess, A., Maran, S.~P., Mushotzky, R.~F., \& Wu, C.-C.\ 
1999, \apj, 516, 750 
\bibitem[Elvis et al.(1989)]{elvis89} Elvis, M., Wilkes, B.~J., 
\& Lockman, F.~J.\ 1989, \aj, 97, 777 
\bibitem[Elvis(2000)]{elvis} Elvis, M.\ 2000, \apj, 545, 63 
\bibitem[Ferland (2004)]{fer98} Ferland, G. J. Korista, K.T. Verner, D.A. Ferguson, J.W. Kingdon, J.B. Verner, 
\& E.M. 1998, PASP, 110, 761 

\bibitem[Fox et al.(2004)]{fox04} Fox, A.~J., Savage, B.~D., 
Wakker, B.~P., Richter, P., Sembach, K.~R., \& Tripp, T.~M.\ 2004, \apj, 
602, 738 
\bibitem[Gabel et al.(2005a)]{gabel05a} Gabel, J.~R., et al.\ 
2005, \apj, 631, 741 
\bibitem[Gabel et al. (2005b)]{gabel05} Gabel, J.~R., et al.\ 
2005, \apj, 623, 85, G05 
\bibitem[Gon{\c c}alves et al.(2001)]{2001MNRAS.328..409G} Gon{\c c}alves, 
D.~R., Fria{\c c}a, A.~C.~S., \& Jatenco-Pereira, V.\ 2001, \mnras, 328, 
409 
\bibitem[Grevesse \& Sauval (1998)]{grevesse98} Grevesse, N.~\& 
Sauval, A.~J.\ 1998, Space Science Reviews, 85, 161 
\bibitem[Kaspi et al. (2001)]{kaspi01} Kaspi, S., Brandt, W.N., Netzer, H., et al., 2001, ApJ 554, 216
\bibitem[Kaastra \& Barr(1989)]{kaastrabarr} Kaastra, J.~S., \& 
Barr, P.\ 1989, \aap, 226, 59 
\bibitem[Kaastra et al. (2002)]{kaastra02} Kaastra, J.~S. et al.\ 2002, \aap, 386, 427 
\bibitem[Kaastra et al. (2004)]{jelle04} Kaastra, J.~S., et al.\ 
2004, \aap, 428, 57 
\bibitem[Kinkhabwala et al.(2002)]{kink02} Kinkhabwala, A., et 
al.\ 2002, \apj, 575, 732 
\bibitem[Korista et al.(1997)]{kor97} Korista, K., Baldwin, 
J., Ferland, G., \& Verner, D.\ 1997, \apjs, 108, 401 
\bibitem[Korista \& Goad (2000)]{korista00} Korista, K.~T., \& Goad, M.~R.\ 2000, \apj, 536, 284 
\bibitem[Kriss(2002)]{kriss02} Kriss, G.~A.\ 2002, ASP 
Conf.~Ser.~255: Mass Outflow in Active Galactic Nuclei: New Perspectives, 
255, 69 
\bibitem[Krolik \& Kriss(1995)]{kriss95} Krolik, J.~H., \& 
Kriss, G.~A.\ 1995, \apj, 447, 512 
\bibitem[Krolik \& Kriss(2001)]{kriss01} Krolik, J.~H., \& 
Kriss, G.~A.\ 2001, \apj, 561, 684 
\bibitem[Krongold et al.(2003)]{k03} Krongold, Y., 
Nicastro, F., Brickhouse, N.~S., Elvis, M., Liedahl, D.~A., \& Mathur, S.\ 
2003, \apj, 597, 832 
\bibitem[Krongold et al. (2005)]{k05} Krongold, Y., 
Nicastro, F., Brickhouse, N.~S., Elvis, M., \& Mathur, S.\ 2005, \apj, 622, 
842 
\bibitem[Lee et al.(2001)]{lee} Lee, J.~C., Ogle, P.~M., 
Canizares, C.~R., Marshall, H.~L., Schulz, N.~S., Morales, R., Fabian, 
A.~C., \& Iwasawa, K.\ 2001, \apjl, 554, L13 
\bibitem[Laor (1991)]{laor} Laor, A., 1991, ApJ 376, 90
\bibitem[Netzer et al.(2002)]{netzer02} Netzer, H., Chelouche, 
D., George, I.~M., Turner, T.~J., Crenshaw, D.~M., Kraemer, S.~B., \& 
Nandra, K.\ 2002, \apj, 571, 256 
\bibitem[Netzer et al. (2003)]{netzer03} Netzer, H., et al.\ 
2003, \apj, 599, 933 
\bibitem[Nicastro et al.(1999)]{nicastro99} Nicastro, F., Fiore, 
F., \& Matt, G.\ 1999, \apj, 517, 108 
\bibitem[Nicastro et al.(2002)]{nicastro02} Nicastro, F., et al.\ 
2002, \apj, 573, 157 
\bibitem[Ogle et al. (2004)]{ogle04} Ogle, P.~M., Mason, K.~O., 
Page, M.~J., Salvi, N.~J., Cordova, F.~A., McHardy, I.~M., \& Priedhorsky, 
W.~C.\ 2004, \apj, 606, 151 
\bibitem[Peterson(1993)]{peterson93} Peterson, B.~M.\ 1993, \pasp, 
105, 247 
\bibitem[Porquet \& Dubau(2000)]{delphine} Porquet, D., \& Dubau, J.\ 2000, \aaps, 143, 495 
\bibitem[Pounds et al.(2004)]{pounds} Pounds, K.~A., Reeves, 
J.~N., King, A.~R., \& Page, K.~L.\ 2004, \mnras, 350, 10 
\bibitem[Proga(2003)]{proga} Proga, D.\ 2003, \apj, 585, 406 
\bibitem[Sako et al.(2000)]{sako00} Sako, M., Kahn, S.~M., Paerels, F., \& Liedahl, D.~A.\ 2000, \apjl, 543, L115 
\bibitem[Sako et al.(2003)]{sako} Sako, M., et al.\ 2003, \apj, 596, 114 
\bibitem[Savage et al.(2003)]{savage} Savage, B.~D., et al.\ 
2003, \apjs, 146, 125 
\bibitem[Scott et al.(2004)]{scott04} Scott, J.~E., et al.\ 
2004, \apjs, 152, 1 
\bibitem[Sembach et al.(2003)]{sembach} Sembach, K.~R., et al.\ 
2003, \apjs, 146, 165 
\bibitem[Steenbrugge et al. (2005)]{steen05} Steenbrugge, K.~C., 
et al.\ 2005, \aap, 434, 569 
\bibitem[Stirpe et al.(1994)]{stirpe} Stirpe, G.~M., et al.\ 
1994, \aap, 285, 857 
\bibitem[Turner et al.(2005)]{turner} Turner, T.~J., Kraemer, 
S.~B., George, I.~M., Reeves, J.~N., \& Bottorff, M.~C.\ 2005, \apj, 618, 
155 
\bibitem[Wandel et al. (1999)]{1999ApJ...526..579W} Wandel, A., Peterson, 
B.~M., \& Malkan, M.~A.\ 1999, \apj, 526, 579 
\bibitem[Wang et al.(2005)]{wang} Wang, Q.~D., et al.\ 2005, astro-ph/0508661 
\bibitem[Weaver et al. (1995)]{1995ApJ...447..121W} Weaver, K.~A., Arnaud, 
K.~A, \& Muschotzky, R.~F.\ 1995, \apj, 447, 121 

\bibitem[Weaver et al. (2001)]{2001ApJ...550..261W} Weaver, K.~A., Gelbord, 
J., \& Yaqoob, T.\ 2001, \apj, 550, 261 
\bibitem[Williams et al.(2006)]{will05} Williams, R.~J., 
Mathur, S., \& Nicastro, F.\ 2006, \apj, 645, 179 



\end{thebibliography}
\end{document}